\documentclass{aa}
\newcommand{\varcsec}{^{\prime\prime}}
\newcommand{\arcsecc}{.\hspace{-0.9mm}'\!\hskip0.4pt'\hspace{-0.2mm}}
\newcommand\ionn[2]{#1$\;${\scshape{#2}}}%
\usepackage{amssymb}
\usepackage{textcomp}
\usepackage{graphicx}
\usepackage{epstopdf}
\usepackage{natbib}
\usepackage{float}
\usepackage{longtable}
\usepackage{microtype}
\usepackage{color} 
\usepackage{xfrac} 
\usepackage{inputenc}
\usepackage{fontenc}
\usepackage{mathtools}
\usepackage{hyphenat}
\usepackage[bookmarks = true,colorlinks = true,linkcolor = blue,urlcolor = blue,citecolor = blue,bookmarks = true,linktocpage = true,hyperindex = true,breaklinks]{hyperref}
\DeclarePairedDelimiter\abs{\lvert}{\rvert}%
\bibpunct{(}{)}{;}{a}{}{,}

\usepackage{txfonts}
\begin{document}
   \title{Vertical magnetic field gradient in the photospheric layers of sunspots}
   
   \subtitle{}

   \author{Jayant Joshi\inst{\ref{inst1},\ref{inst2}}
           \and Andreas Lagg \inst{\ref{inst1}}
           \and Johann Hirzberger\inst{\ref{inst1}}
           \and Sami K. Solanki\inst{\ref{inst1},\ref{inst3}}
           \and Sanjiv K. Tiwari \inst{\ref{inst1},\ref{inst4}}}
   \institute{Max-Planck-Institut f\"{u}r Sonnensystemforschung, Justus-von-Liebig-Weg 3,
              37077, G\"{o}ttingen, Germany\label{inst1} 
              \and Institute for Solar Physics, Department of Astronomy, Stockholm University, AlbaNova University Centre,
              SE-106 91 Stockholm, Sweden\label{inst2}
              \and School of Space Research, Kyung Hee University, Yongin, Gyeonggi Do, 446-701, 
              Republic of Korea\label{inst3}
              \and NASA Marshall Space Flight Center, Mail Code ZP 13, Huntsville, AL 35812, USA\label{inst4}\\
              \email{jayant.joshi@astro.su.se} 
              }
   \date{Received; accepted}

 
  \abstract
  { 
   }  
  {We investigate the vertical gradient of the magnetic field of sunspots in the photospheric layer.
     }
  {Independent observations were obtained with the Solar Optical Telescope/Spectropolarimeter 
     (SOT/SP) onboard the Hinode spacecraft and with the Tenrife Infrared Polarimeter-2 (TIP-2) 
     mounted at the German Vacuum Tower Telescope (VTT). We apply 
     state-of-the-art inversion techniques to both data sets to 
     retrieve the magnetic field and the corresponding vertical 
     gradient along with other atmospheric parameters in the solar 
     photosphere. }
  {In the sunspot penumbrae we detected patches of 
     negative vertical gradients of the magnetic field strength, 
     i.e.,  the magnetic field strength decreases with optical depth in 
     the photosphere. The negative gradient patches are located in the inner and partly in the middle
     penumbrae in both data sets. From the SOT/SP observations we found that 
     the negative gradient patches are restricted mainly to the 
     deep photospheric layers and are concentrated near the edges 
     of the penumbral filaments. MHD simulations also show negative 
     gradients in the inner penumbrae, also at the locations of filaments. 
     Both in the observations and simulation negative gradients of the magnetic field 
     vs. optical depth dominate at some radial distances in the penumbra. 
     The negative gradient with respect to optical depth in the inner penumbrae persists even after averaging 
     in the azimuthal direction, both in the observations and, to a lesser extent, also in MHD simulations.
     If the gradients in the MHD simulations are determined with respect to geometrical height, 
     then the azimuthal averages are always positive within the sunspot (above $\log \tau = 0$), corresponding 
     to magnetic field increasing with depth, as generally expected.}
   {We interpret the observed localized presence of negative vertical gradient of the magnetic field
    strength in the observations as a consequence of stronger field from spines expanding with height and closing above the weaker 
    field inter-spines. The presence of the negative gradients with respect to optical depth after azimuthal
    averaging can be explained by two different mechanisms: 
     the high corrugation of equal optical depth surfaces and 
     the cancellation of polarized signal due to the 
     presence of unresolved opposite polarity patches in the deeper 
     layers of the penumbra.}

   \keywords{Sun: magnetic field - Sun: activity - Polarimetry} 

   \titlerunning{Vertical magnetic field gradient of sunspots}
   \authorrunning{Jayant Joshi et. al.}

   \maketitle
%

\section{Introduction}

The vertical gradient of the magnetic field vector in sunspot
photospheres has been studied extensively in the past \citep[for 
reviews see, e.g.,][]{Solanki_2003,Borrero_2011}. Depending on the diagnostic
tools, a wide range of values from 1.0\, to 4.0\,G\,km$^{-1}$ 
for the vertical gradient of the magnetic field strength in sunspot 
umbrae have been reported \citep{Westendrop_2001a,Mathew_2003,Orozco_2005,Sanchez_cuberes_2005,
Balthasar_2008,Borrero_2011}.
Positive values indicate an increase of the magnetic field strength with geometrical depth. Studies that
display the vertical field gradient as a function of the normalized
radius of the sunspot also differ
significantly. \citet{Westendrop_2001a}, \citet{Borrero_2011} and
\citet{Orozco_2005} interpret the observations to say that the sunspot magnetic field forms
canopy-like structures in the middle and outer penumbrae. In contrast,
\citet{Mathew_2003,Sanchez_cuberes_2005,Balthasar_2008} merely conclude that the
magnetic field increases with depth everywhere in the sunspot. \citet{Tiwari_2015} also
do not find evidence for such a canopy-like structure in the middle and outer penumbra, although
they report on a reversed magnetic field gradient in the inner penumbra 
(i.e. field strength decreasing with depth), and field canopy starting 
just outside the sunspot penumbra.
Consequently there is a need to revisit the question 
of the behavior of the vertical field gradient in sunspots.  

In the present paper we study the radial dependence of the vertical
gradient of the magnetic field strength of a sunspot observed on two 
different days by Tenerife Infrared Polarimeter-2 \citep[TIP-2;][]{Collados_2007}
mounted at the Vacuum Tower Telescope (VTT)
and of another sunspot observed by the Spectropolarimeter of the Solar Optical Telescope 
\citep[SOT/SP;][]{Tsuneta_2008,Shimizu_2008,Suematsu_2008,Ichimoto_2008}. 
To retrieve the atmospheric parameters from VTT/TIP-2 data we carry out inversions of Stokes profiles
of the \ionn{Si}{i}\,10827.1\,\AA\, and \ionn{Ca}{i}\,10833.4\,\AA\, spectral lines 
together. We also use spatially
coupled inversions to the Hinode SOT/SP observations (i.e.,  the \ionn{Fe}{i}\,6301.5\,\AA\,
and \ionn{Fe}{i}\,6302.5\,\AA\, lines) to determine the vertical field gradients.
We compare the results from the different data sets with each other 
and with a 3D MHD simulation of a sunspot by \citet{Rempel_2009a}.

\section{VTT/TIP-2 observations and analysis} \label{4_tip}

\subsection{Observations} \label{4_tip_obs}

We observed the leading sunspot in the active region NOAA~11124 on 14
and 16 November, 2010.  Spectro-polarimetric observations were carried out at
the VTT/TIP-2. The spectral domain of
the observations ranges from 10825\,\AA\, to 10835\,\AA\ and the
spectral sampling corresponds to 10\,m\AA\, per pixel. The diffraction
limit of the VTT at 10830\,\AA\ is approximately $0\arcsecc 40$, but due to 
moderate seeing conditions the spatial resolution of our data is 
reduced to $\sim 1\arcsecc 0$. Scans of the $\sim 36 \arcsec$ wide field-of-view (FOV) for
full Stokes vector measurements require $\sim 20$ minutes. The 
step-size for scanning was set to $0\arcsecc 36$ and the 
pixel size in the slit direction was $0\arcsecc 16$. In order to improve 
the signal-to-noise ratio we binned four pixels in the direction 
of dispersion and two pixels in slit direction. The heliocentric 
coordinates of the centers of the observed FOVs were (12\textdegree N, 10\textdegree W, $\mu= 0.96$) 
on 14 November, 2010 and (14\textdegree N, 32\textdegree W, $\mu= 0.84$) 
on 16 November, 2010.

\begin{figure}
\centering
   \includegraphics[width=0.48\textwidth]{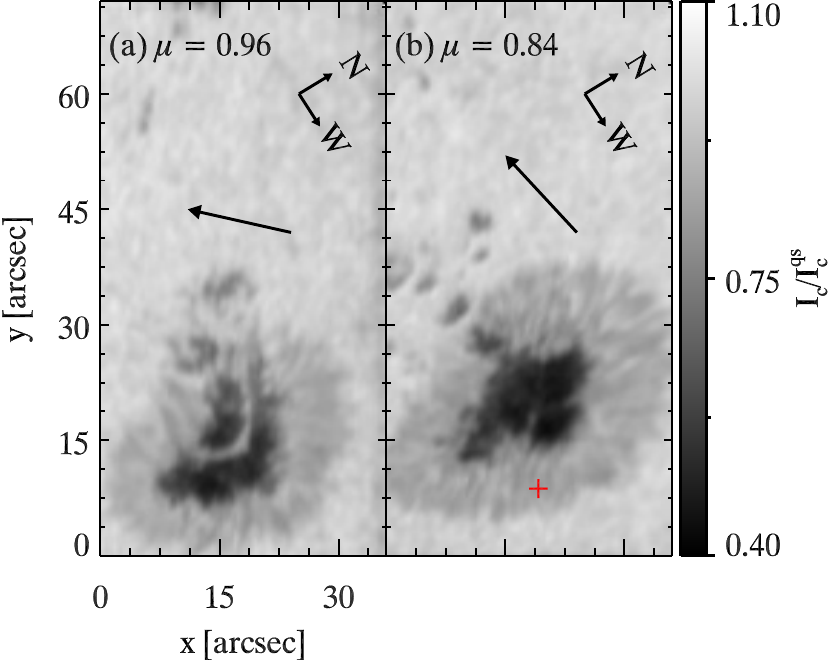}
      \caption{(a) Normalized  continuum intensity  $I_{c}/I_{c}^{qs}$ map of the observed sunspot in
        AR NOAA 11124 observed on 14 November, 2010, obtained
        from VTT/TIP-2 data. $I_{c}$ denotes the continuum intensity and $I_{c}^{qs}$ is the averaged quiet Sun
        continuum intensity. (b) $I_{c}/I_{c}^{qs}$ map of the same sunspot 
        observed on 16 November, 2010. Arrows in both maps indicate 
        the direction to disk center. Solar north and west directions are marked by arrows in the \textit{upper-right}
        part of panels.
        }
\label{cont}
\end{figure}

Panels (a) and (b) of Fig.~\ref{cont} show the sunspot on
14 and 16 November, 2010, respectively, as observed in the continuum
intensity at 10833\,\AA\,. The sunspot shows a light-bridge on 14 November,  which
disappeared after two days. In general, the sunspot has grown and matured in these two days,
being more symmetric and simpler on 16 November, 2010.

The collected data have been treated by applying the standard 
data reduction steps \citep{Collados_1999,Collados_2003}. The continuum level was corrected 
using a Fourier Transform
Spectrometer (FTS) spectrum \citep{Livingston_1991,Wallace_1993}. 

\subsection{SPINOR inversion of the \ionn{Si}{i} and \ionn{Ca}{i} lines} \label{4_tip_inv}

To analyze the photospheric properties of the sunspot's magnetic field we 
apply inversions of Stokes profiles of the \ionn{Si}{i}\,10827.1\,\AA\, and
\ionn{Ca}{i}\,10833.4\,\AA\, lines to satisfy the radiative transfer
equation (RTE) under the assumption of local thermodynamic equilibrium
(LTE). 

\begin{figure}
\centering
   \includegraphics[width=0.48\textwidth]{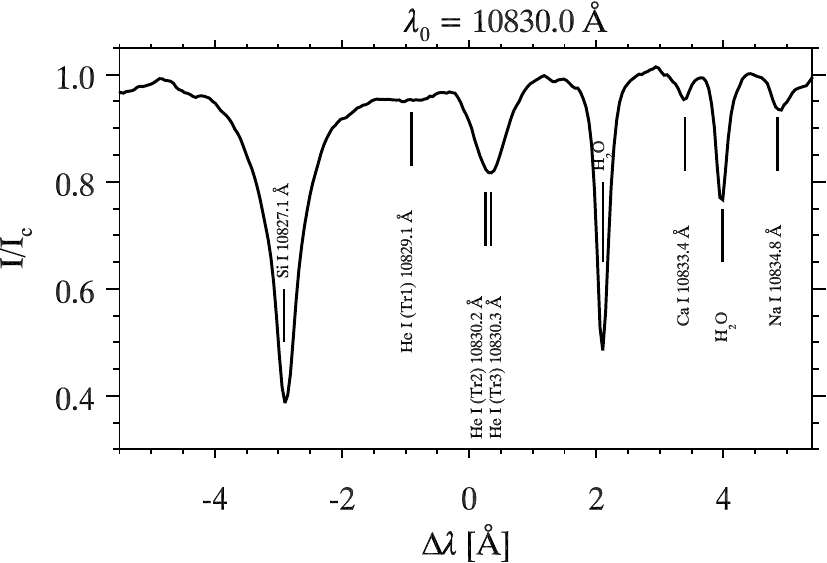}
       \caption{Averaged quiet Sun spectrum obtained with the VTT/TIP-2 on 14 November, 2010.
               $\Delta\lambda=\lambda-\lambda_{o}$, where $\lambda_{o}=10830.0$\,\AA.}
\label{spec}
\end{figure}

\begin{figure}
\centering
   \includegraphics[width=0.48\textwidth]{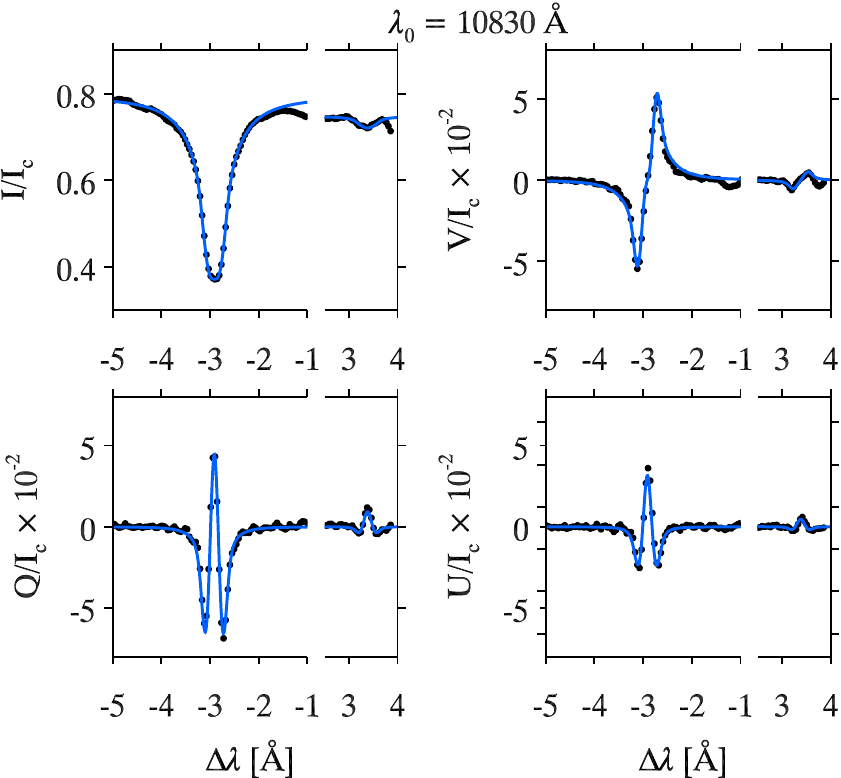}
      \caption{Best fit SPINOR inversions of sample Stokes profiles observed by
               VTT/TIP-2: \textit{Dotted black} curves represent observed Stokes profiles
               and \textit{solid blue} curves represent best fits. The spatial position of 
               the set of Stokes profiles shown here is marked by a \textit{red} plus 
               sign in Fig.~\ref{cont}(b).
              }
\label{fit}
\end{figure}

\begin{figure}
\centering
   \includegraphics[width=0.48\textwidth]{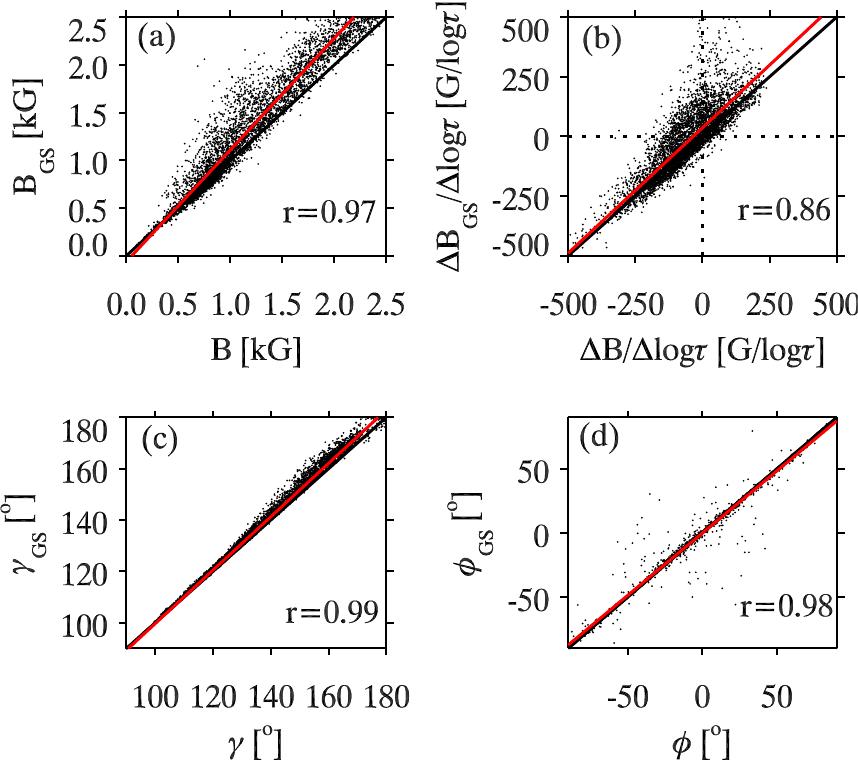}
      \caption{Scatter plots of the magnetic field components obtained from SPINOR inversions without considering straylight vs.
               those with straylight taken into account. The plotted data were obtained on 14 November, 2010,
               with VTT/TIP-2. Panels (a), (b), (c) and (d) show $B$, 
               $\Delta B/\Delta \log \tau$, $\gamma$ and $\phi$, respectively. The \textit{red}
               line in all panels represents the linear regression. The correlation coefficient,
               $r$, is shown at the \textit{lower right} corner of each panel.
               }
\label{GS}            
\end{figure}

\begin{table}
\caption{Atmospheric and atomic parameters of the lines 
used for inversions of the VTT/TIP-2 data.\label{tab5.1}}

\begin{center}
\begin{tabular}{ c c c c c}

\hline\hline
Line & wavelength [\AA] & $\log(gf)$ & Abundance & $g_{\rm{eff}}$\\
\hline
\ionn{Si}{i} & 10827.1 & 0.210 & 7.549 & 1.5\\
\ionn{Ca}{i} & 10833.4 & 0.058 & 6.360 & 1.5\\
\hline

\end{tabular}
\end{center}
\end{table}

The combination of the strong \ionn{Si}{i}\, line and the weak \ionn{Ca}{i}\, line
puts more constraints on the free
parameters in the inversion process and gives more reliable atmospheres than 
inversion of a single-line, especially when investigating the height stratification of the atmosphere. 
The employed line parameters of both lines are given in Table~\ref{tab5.1}. 
The abundance of Si and $\log(gf)$ of the \ionn{Si}{i}\,10827.1\,\AA\, line is
inferred form \citet{Shchukina_2012} and \citet{Shi_2008}, respectively.
The abundance of Ca is inferred from \citet{Grevesse_1998}. 
We estimated the value of $\log(gf)$ for the \ionn{Ca}{i}\,10833.4\,\AA\, line
by fitting the quiet Sun profile observed with the FTS \citep{Livingston_1991,Wallace_1993}
using the HSRASP atmosphere \citep{chapman_1979}. The quiet Sun spectrum obtained 
by averaging 255 quiet sun pixels observed on 14 November 2010 is displayed in Fig.~\ref{spec}. 
To avoid undue influence of the weak \ionn{Ti}{i}\, blend at 10833.66\,\AA\, we have 
restricted the inversion of the \ionn{Ca}{i}\, line to the wavelength range 10833.08 to 10833.52 \AA.

\citet{Bard_2008} have shown that the line core of \ionn{Si}{i}\,10827.1\,\AA\, is 
affected by non-local thermodynamic equilibrium (NLTE) conditions. They 
demonstrated that the line core intensity of \ionn{Si}{i}\,10827.1\,\AA\, under NLTE 
conditions is lower than the line core intensity under LTE conditions. 
\citet{Kuckein_2012a} studied the affect of neglecting NLTE effects
on atmospheric parameters inferred from inversions. They found that the most affected parameter 
is the temperature while the influence on the magnetic field vectors and 
velocities are negligible. Since in this study the temperature stratification 
is of minor relevance, we use the simpler LTE approach for our inversions.

To retrieve the atmospheric parameters in the photosphere, the SPINOR
inversion code \citep{Frutiger_1999,Frutiger_2000, Vannoort_2012} was used,
which is based on the STOPRO routines \citep{Solanki_1987_PhDT}. The
initial model atmosphere consists of a magnetic field strength,
\textit{B}, which varies linearly with the logarithm to the power of 10 of the optical depth ($\log \tau$).
In other words, it is similar to an atmosphere with two nodes for the magnetic 
field strength if the two nodes were the lower most ($\log \tau = 0.0$) and upper 
most ($\log \tau = -2.3$) nodes in our inversion. Hence, the magnetic field has two degrees of freedom.
Three nodes for the line-of-sight (LOS) velocity, $v_{\rm{los}}$, and the temperature, $T$, have 
been used. The other atmospheric parameters such as inclination of
the magnetic field relative to LOS,
$\gamma$, its azimuth direction, $\phi$, and the micro-turbulent velocity, 
$v_{\rm{mic}}$, have been forced to be constant with height. The three nodes 
used are at $\log(~\tau_{\rm{630}})= 0.0, -0.7$ and $-2.3$,
where $\tau_{\rm{630}}$ corresponds to the optical depth at 630\,nm\footnote{
This reference wavelength for optical depth $\tau$ was chosen to be 630\,nm 
throughout this paper, to facilitate comparison with Hinode SOT/SP 
observations.}. Selection of nodes was done by looking at magnetic field response functions, $\rm{RF}_\mathnormal{B}$. 
\citet{Joshi_2014} presented $\rm{RF}_\mathnormal{B}$ of the \ionn{Si}{i}\, and \ionn{Ca}{i}\ lines 
for an umbral and different penumbral atmospheres inferred from the MHD 
simulation of a sunspot by \citet{Rempel_2009a} \citep[see Fig.~4.3 and 4.5 of chapter~4 of][]{Joshi_2014}.
They showed that these lines have large values of
$\rm{RF}_\mathnormal{B}$ at $\log \tau = -0.7$ and $-2.3$  and non-zero values at  $\log \tau= 0.0$.
An example of observed Stokes profiles and best-fit Stokes 
profiles through SPINOR inversions is presented in Fig.~\ref{fit}. In general, the match between synthesized 
and observed profiles is very good.

\begin{figure*}
\sidecaption
\centering
   \includegraphics[width=0.62\textwidth]{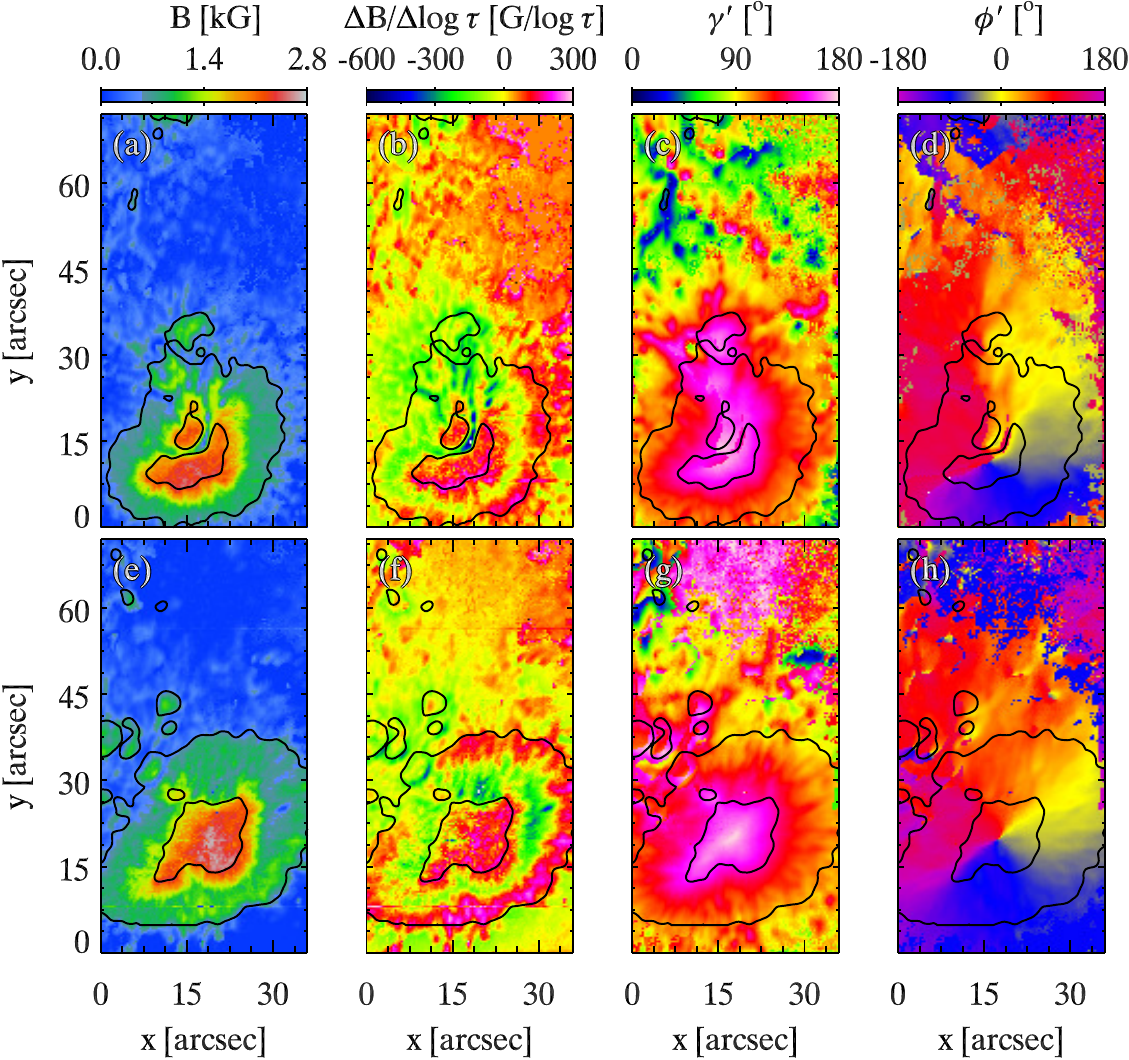}
      \caption{Magnetic field vector obtained from the observations recorded
               on 14 and 16 November, 2010 with VTT/TIP-2. First and Second row of panels
               display the magnetic vector obtained on 14 and 16 November 2010, respectively. 
               (a)/(e) Magnetic field strength $B$, (b)/(f) gradient of the magnetic field strength with respect to 
               $\log \tau$, $(\Delta B/\Delta \log \tau)_{\rm{0.0,-2.3}}$, (c)/(g) inclination angle, $\gamma\prime$, 
               of the magnetic field vector and (d)/(h) azimuth direction, $\phi\prime$, of the magnetic 
               field vector. Both $\gamma\prime$ and $\phi\prime$ are in the solar reference frame.
               The zero value of $\phi\prime$ corresponds to the normal to the line of symmetry 
               (line connecting the geometrical center of sunspot to the solar \textbf{disk center}).  
              }
\label{mag14}
\end{figure*}

\subsubsection{Influence of straylight} \label{4_tip_stray}


To examine the effect of straylight on the retrieved magnetic 
field vector and on the vertical gradient of $B$, we fitted the same atmospheric 
model with a second atmospheric component, representing the contribution of 
global straylight. By global 
straylight we mean that the straylight function does not vary spatially 
in the observed FOV. It is assumed to originate from the   
the broad wings of the point-spread function, typical for seeing-affected ground-based observations, and 
therefore resembling the shape of the averaged quiet Sun Stokes $I$ profile. 
The straylight contribution is denoted by $\alpha$ in 
Eqn.~\ref{eqn_5.1}:

\begin{equation}  \label{eqn_5.1}
I_{O}=(1-\alpha)I_{L}+\alpha\,I_{GS}.
\end{equation}

\noindent
Here $I_{\rm{o}}$ is the observed Stokes $I$, $I_{\rm{L}}$ represents Stokes $I$
of the local component and $I_{\rm{GS}}$ is Stokes $I$ 
of the straylight component. $\alpha$ is an additional free parameter in 
the inversions. We are aware that this is a simplified approach that neglects that
the absolute amount of straylight might decrease with distance from the quiet Sun and that 
in the penumbra polarized straylight, probably, is important as well. This approach is 
taken purely for test purposes since the fits of the synthesized Stokes profiles 
to the observed Stokes profiles are very good even without any consideration of straylight.

A comparison of the obtained magnetic field vectors inferred with and without 
straylight is presented in Fig.~\ref{GS}, showing very good correlation. Among all parameters the 
vertical gradient displays the least correlations (86\%). Although $B$ 
is somewhat underestimated (by 6\% in the umbra) when straylight is not considered, we see negative
and positive vertical gradients of $B$ in both approaches to the inversion. This test suggests that 
neglecting straylight does not effect the qualitative structure of the vertical gradient of 
the magnetic field in the sunspot, although there are some quantitative differences. 
For the analysis presented in this section we therefore use 
the parameters retrieved from inversions neglecting the straylight contribution.

For the Hinode/SOT data the spatial PSF is constant and well known and therefore the 
influence of this spatially inhomogeneous straylight is taken into 
account (for details see \citet{Vannoort_2012}, \citet{Vannoort_2013} and Section~\ref{4_sot}).

\begin{figure*}

\centering
   \includegraphics[width=0.8\textwidth]{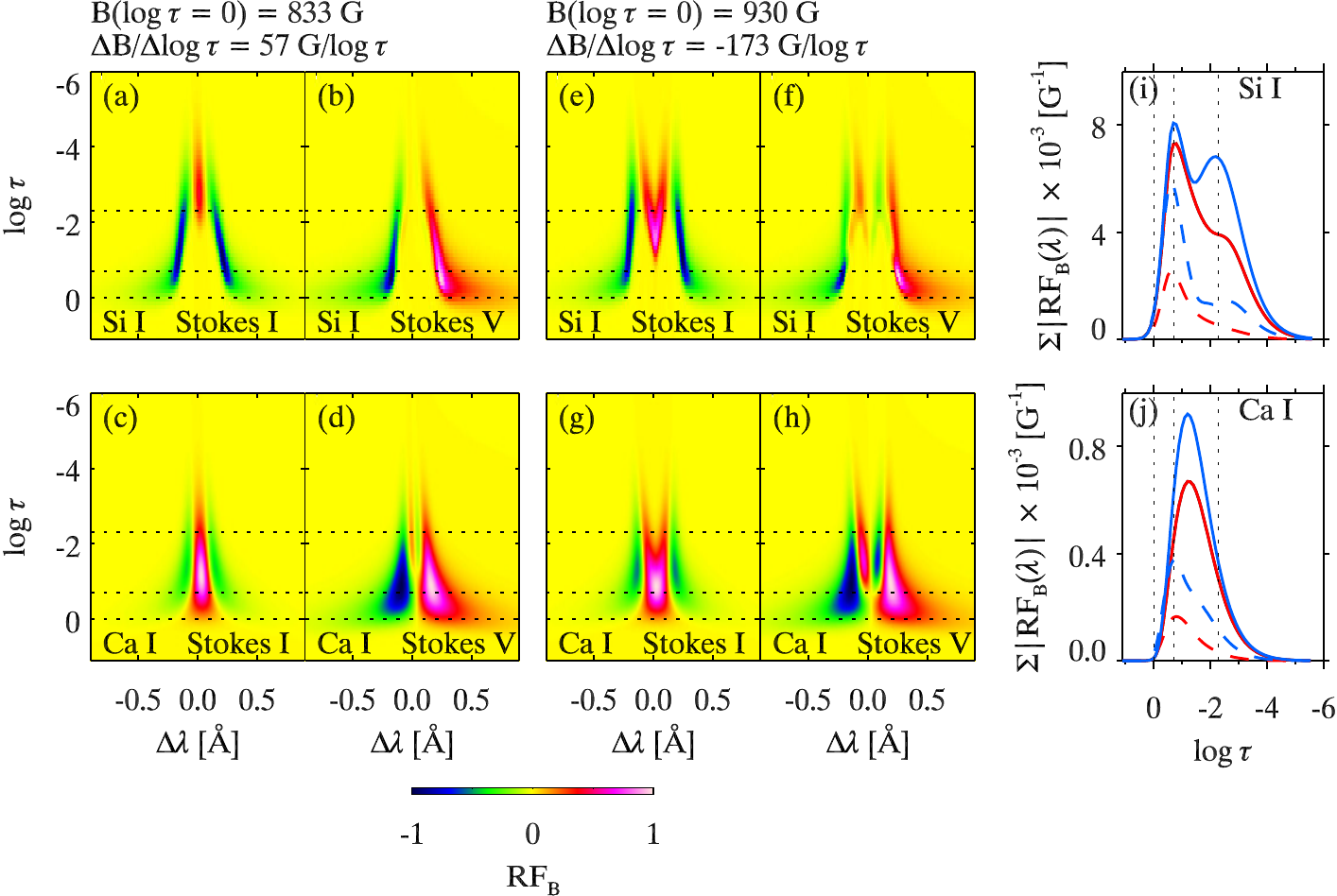}
      \caption{Magnetic field response functions, $\rm{RF}_\mathnormal{B}$, of the \ionn{Si}{i}\,10827.1\,\AA\, and 
                \ionn{Ca}{i}\,10833.4\,\AA\, lines. Panels (a) and (b) display $\rm{RF}_\mathnormal{B}$ of Stokes $I$
                and Stokes $V$ of the \ionn{Si}{i}\,10827.1\,\AA\, line respectively, for a penumbral atmosphere where 
                $(\Delta B/\Delta \log \tau)$ is positive. For the same atmosphere, $\rm{RF}_\mathnormal{B}$ of \ionn{Ca}{i}\,10833.4\,\AA\, are shown 
                in panels (c) and (d). Panels (e)-(h) are same as (a)-(d) but represent 
                $\rm{RF}_\mathnormal{B}$ of a penumbral atmosphere where $(\Delta B/\Delta \log \tau)$ is negative. Displayed
                $\rm{RF}_\mathnormal{B}$ are normalized to the maximum value. 
                The absolute values of $\rm{RF}_\mathnormal{B}$ integrated along the wavelength range  
                $\Delta\lambda=\pm$ 1.0\,\AA\, for the \ionn{Si}{i}\, and \ionn{Ca}{i}\, lines are shown in panels 
                (i) and (j), respectively. \textit{Red} and \textit{blue} curves correspond to the positive and negative $(\Delta B/\Delta \log \tau)$ 
                respectively. \textit{Solid} curves are for Stokes $I$ and \textit{Dashed} curves are for Stokes $V$. 
                \textit{Dotted} horizontal or vertical lines in the panels indicate the $\log \tau$ nodes used in the inversions.
               }
\label{brf}
\end{figure*}

\begin{figure}
\centering
   \includegraphics[width=0.48\textwidth]{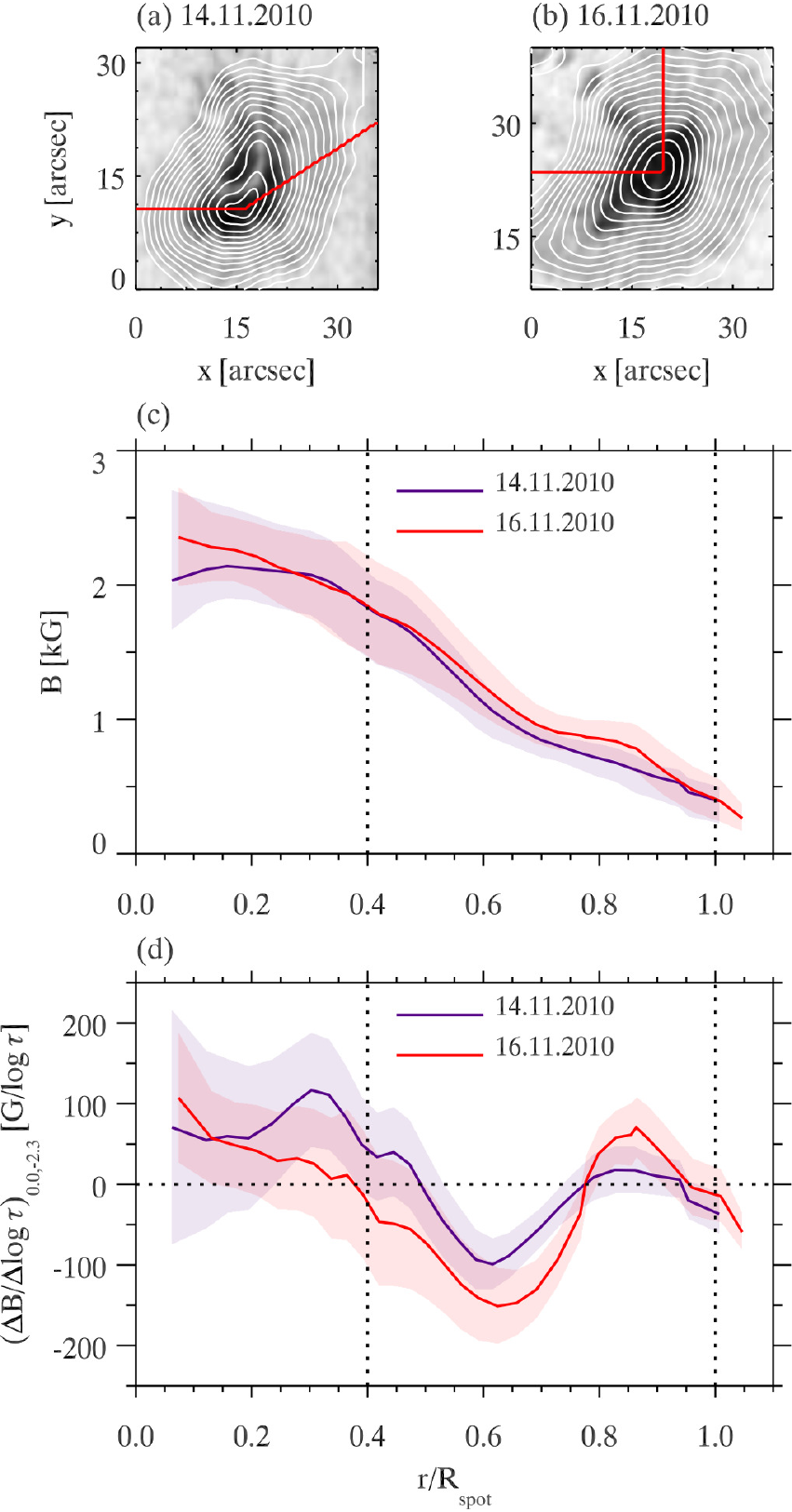}
      \caption{Continuum intensity maps from 14 November, 2010 (panel (a)) 
               and 16 November, 2010 (panel (b)). The contours are used to 
               calculate azimuthal averages. 
               The areas \textbf{above} the \textit{red} lines were neglected because 
               of the complex structure of the sunspot there.
               Azimuthally averaged parameters: Panel (c) shows the radial dependence of \textit{B} of the sunspot 
               observed on 14 November, 2010 (\textit{violet} curve) and 16 November, 2010
               (\textit{red} curve) with VTT/TIP-2. Panel (d) represents the vertical gradient 
               of the magnetic field strength  $(\Delta B/\Delta \log \tau)_{0.0,-2.3}$ as a function of 
               $r/R_{\rm{spot}}$.\,\textit{Shaded} areas represent standard deviations.
               \textit{Dotted} vertical lines in panels (c) and (d) indicate
               the averaged location of the umbra-penumbra boundary and the outer boundary of the sunspot.
               }
\label{radial_tmp}
\end{figure}

\subsection{Inversion results} \label{4_tip_res}

The magnetic field parameters retrieved from inversions of the observations 
from 14 November, 2010 are presented in the upper row of Fig.~\ref{mag14}. Panels (a) and (b) 
display maps of \textit{B} and and its linear gradient, $(\Delta B/\Delta \log \tau)_{\rm{0.0,-2.3}}$ 
of the observed FOV, respectively. The definition of $(\Delta B/\Delta \log \tau)$ is given by

\begin{equation} \label{eqn_5.2}
{\left(\frac{\Delta B}{\Delta \log \tau}\right)_{\rm{a,b}}=\frac{(\Delta B)_{\rm{a,b}}}{(\Delta\log \tau)_{\rm{a,b}}}=\frac{B(b)-B(a)}{b-a}} \,,
\end{equation}
\noindent
where a and b denote lower and upper $\log \tau$ surfaces, respectively. Inclination and azimuth angles of the
magnetic field vectors,\,\textbf{\textit{B}}, are presented in panels (c) and (d) of Fig.~\ref{mag14}.
The magnetic field vectors presented here are projected to disk center 
coordinates using the transformation matrix given by \citet{Wilkinson_1989}. The 
180\textdegree\, ambiguity in azimuth direction  
is resolved by the \textgravedbl acute angle\textacutedbl\, 
method \citep{Sakurai_1985,Cuperman_1992}. Inclination and azimuth angle of 
\textbf{\textit{B}} after correction to the solar disk center coordinates
are denoted by $\gamma\prime$ and $\phi\prime$, respectively. 
Magnetic field parameters obtained from the observations 
recorded on 16 November, 2010 are shown in the lower row of Fig.~\ref{mag14}.
The umbral and penumbral boundaries shown here as well as those for the SOT/SP observations 
and MHD simulation (presented in Section~\ref{4_sot} and Section~\ref{4_mhd} respectively)
are determined by the method of cumulative histogram of the continuum intensity as demonstrated by \citet{Mathew_2007}. 

Whereas on 14 November the maximum of \textit{B} amounts to $\sim 2500$\,G 
and appears in the darkest part of the umbra it increases up to $\sim 2800$\,G 
two days later. Maps of $(\Delta B/\Delta \log \tau)_{\rm{0.0,-2.3}}$ show 
consistent results on both days:  
in the umbra $(\Delta B/\Delta \log \tau)_{\rm{0.0,-2.3}}$ is positive, i.e., \textit{B}
increases with optical depth. 
In the undisturbed part of the inner and middle penumbra the gradient maps show a ring-like 
structure where $(\Delta B/\Delta \log \tau)_{\rm{0.0,-2.3}}$ is negative, i.e., $B$ decreases with optical depth. 
In the outer penumbra $(\Delta B/\Delta \log \tau)_{\rm{0.0,-2.3}}$ is positive and just outside 
the visible boundary of the sunspot, $(\Delta B/\Delta \log \tau)_{\rm{0.0,-2.3}}$ is negative again.

The magnetic field response functions, $\rm{RF}_\mathnormal{B}$, of Stokes $I$ and $V$ of the spectral
line diagnosed here show that $\log \tau$ nodes used in the inversions roughly cover the range of heights
to which the lines are sensitive. Fig.~\ref{brf} displays $\rm{RF}_\mathnormal{B}$ for two different locations
in the penumbra with different sign of $(\Delta B/\Delta \log \tau)_{\rm{0.0,-2.3}}$. $\rm{RF}_\mathnormal{B}$
for a penumbral atmosphere where $B_{\rm{\log \tau=0.0}}$ = 833\,G,   
$(\Delta B/\Delta \log \tau)_{\rm{0.0,-2.3}}$ = 57\,G/$\log \tau$ and $T_{\log \tau = 0.0}$ = 6273\,K\, are presented in 
Fig.~\ref{brf}(a)-(d), while  Fig.~\ref{brf}(e)-(h) show $\rm{RF}_\mathnormal{B}$ for an atmosphere where $B_{\rm{\log \tau=0.0}}$ = 930\,G,  
$(\Delta B/\Delta \log \tau)_{\rm{0.0,-2.3}}$ = --173\,G/$\log \tau$ and $T_{\log \tau = 0.0}$ = 6328\,K. Wavelength-integrated $\abs{\rm{RF}_{B}}$ are plotted in 
Fig.~\ref{brf}(i)  and (j), showing that these lines have strong $\rm{RF}_\mathnormal{B}$ at $\log \tau = -0.7$ and $-2.3$ 
and non-zero values at $\log \tau = 0.0 $ irrespective of the nature of $(\Delta B/\Delta \log \tau)_{\rm{0.0,-2.3}}$.
The model atmosphere in our inversions consists of a magnetic field strength which varies linearly with $\log \tau$.

To study the radial dependence of the properties of the sunspot
atmosphere, we use azimuthal averages of all parameters. The azimuthal
averages are computed along 25 isothermal contours obtained from a smoothed
temperature map at $\log \tau = 0.0$ (see panels (a)
and (b) of Fig.~\ref{radial_tmp}). We do not consider the parts of the penumbra 
below the \textit{red} lines shown in Fig.~\ref{radial_tmp}(a) and (b), since the field structure 
there is strongly disturbed by a light bridge and an arch filament system connecting this region 
to points outside the FOV. 

\begin{figure*}
\sidecaption
\centering
   \includegraphics[width=0.74\textwidth]{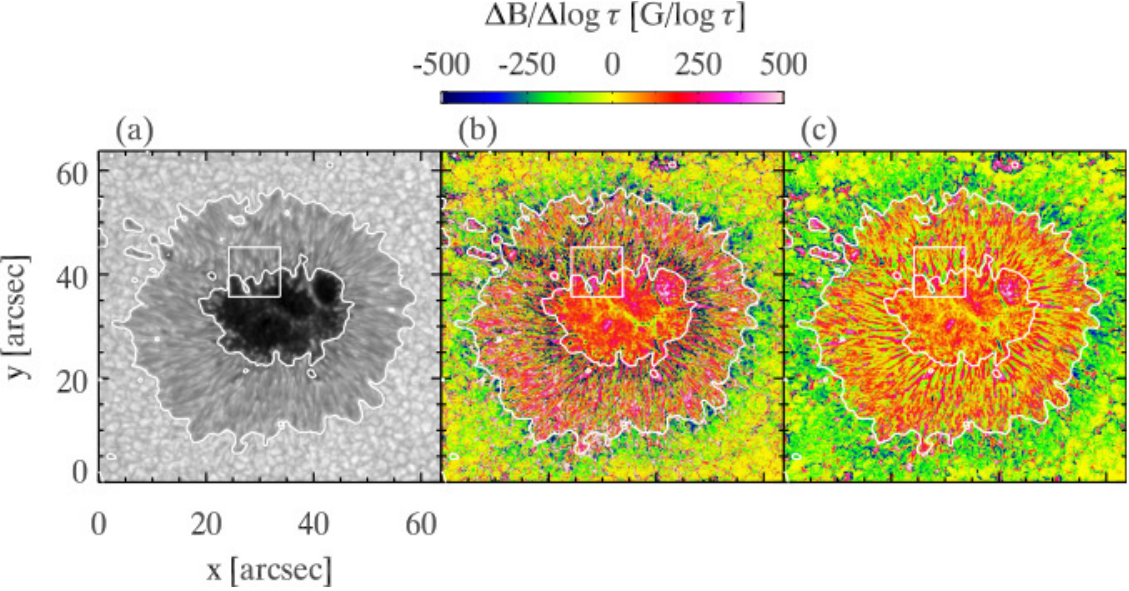}
      \caption{Continuum intensity (panel (a)) and $(\Delta B/\Delta \log \tau)$ maps of 
               a sunspot in active region NOAA 10933, observed with Hinode 
               SOT/SP. Panels (b) and (c) represent
               $(\Delta B/\Delta \log \tau)_{0.0,-0.9}$ and $(\Delta B/\Delta \log \tau)_{-0.9,-2.5}$,
               respectively. Inner 
               and outer \textit{white} contours in all panels represent the umbra-penumbra 
               boundary and the outer boundary of the sunspot, respectively.
               }
\label{skt_maps}
\end{figure*}

\begin{figure*}
\centering
   \includegraphics[width=0.8\textwidth]{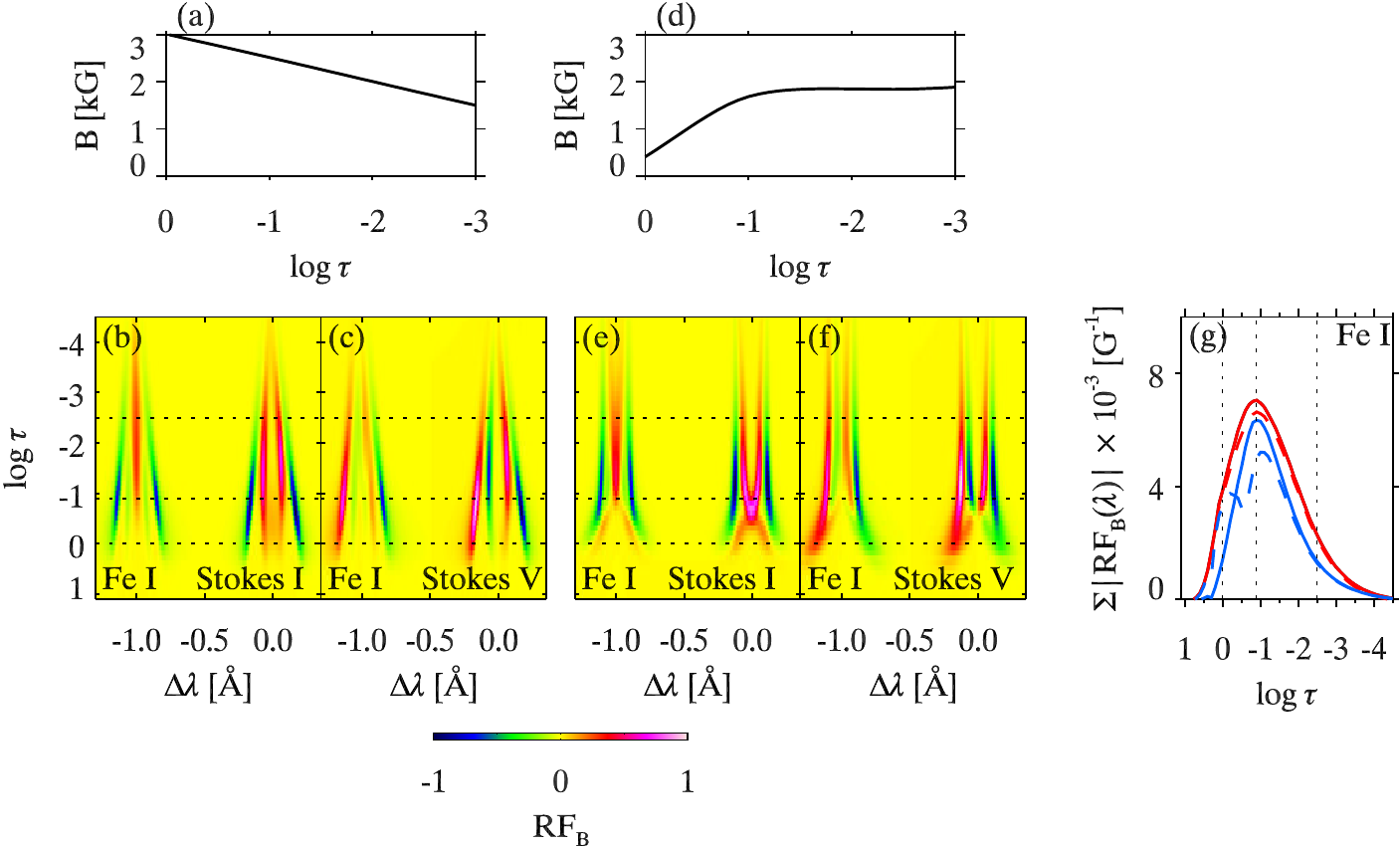}
      \caption{Magnetic field response functions, $\rm{RF}_\mathnormal{B}$, of the  
               \ionn{Fe}{i}\,6301.5\,\AA\, and \ionn{Fe}{i}\,6302.5\,\AA\, lines.
               Panels (b) and (c) show $\rm{RF}_\mathnormal{B}$ for Stokes $I$ and Stokes $V$, respectively, for a penumbral atmosphere
               with the magnetic field strength stratification presented in panel (a). Panels (e) and (f)
               are the same as panels (b) and (c) but correspond to the magnetic field stratification displayed in panel (d).
               The presented $\rm{RF}_\mathnormal{B}$ values are normalized to the maximum value.
               The absolute values of $\rm{RF}_\mathnormal{B}$, integrated over wavelength 
               ($\Delta\lambda=\pm$ 0.35\AA\,) for the \ionn{Fe}{i}\,6302.5\,\AA\, line, are shown in panel 
               (g). \textit{Red} and \textit{blue} curves correspond to the positive and negative $(\Delta B/\Delta \log \tau)$, 
                respectively. \textit{Solid} curves are for Stokes $I$ and \textit{Dashed} curves are for Stokes $V$.
               \textit{Dotted} horizontal lines in  
               panels (b), (c), (e) and (f) and  \textit{Dotted} vertical lines in panel (g) 
               indicate  the $\log \tau$ nodes used in the inversions.
              }
\label{fe_rfb}
\end{figure*}

\begin{figure*}
\sidecaption
\centering
   \includegraphics[width=0.70\textwidth]{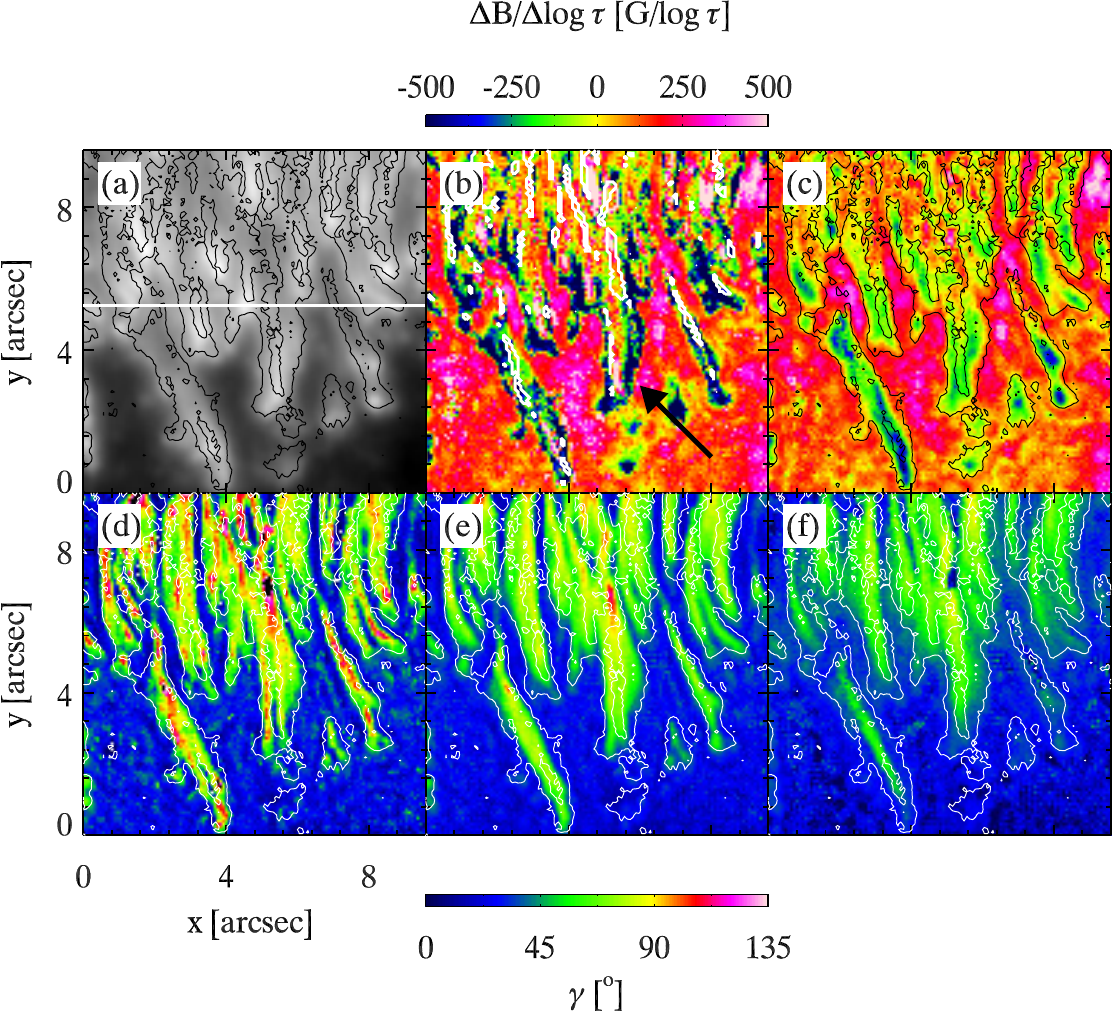}
      \caption{(a)-(c): Same as Fig.~\ref{skt_maps}, but showing only the area within 
               the \textit{white} box in panel (a) of Fig.~\ref{skt_maps}. Additional panels (d), (e) and (f)
               displaying inclination of the magnetic field at $\log \tau = 0.0, -0.9, -2.5$, respectively. 
               Contours in the panels (a), (c), (d), (e) and (f) separate patches of negative and positive
               $(\Delta B/\Delta \log \tau)_{0.0,-0.9}$ and \textit{white} contours in
               the panel (b) indicate areas of polarity opposite to that in the 
               umbra at $\log \tau$ = 0.0. The horizontal \textit{white} line in panel (a) indicates the 
               location of the cut plotted in Fig.~\ref{skt_slit}. The \textit{Black} arrow in panel (b) marks
               an example of a penumbral filament with negative $(\Delta B/\Delta \log \tau)_{0.0,-0.9}$ at 
               its edges.
               }
\label{skt_maps_roi}
\end{figure*}

The radial dependences of $B$ and $(\Delta B/\Delta \log \tau)_{\rm{0.0,-2.3}}$ are displayed in 
Fig.~\ref{radial_tmp}(c) and (d). On 14 (16) November, $(\Delta B/\Delta \log \tau)_{\rm{0.0,-2.3}}$ has a value of 
120 (100)\,G/$\log \tau$ on average in the darkest part of umbra with a maximum 
amplitude of 185 (175)\,G/$\log \tau$. Form these values we estimate the average magnetic field gradient in the 
umbra on a geometrical depth scale is to be $\sim1.3$\,G\,km$^{-1}$. 
We converted $\Delta\log \tau$ to $\Delta d$ assuming hydrostatic equilibrium,
where $d$ is the geometrical depth. In the inner penumbra, i.e., from $r/R_{\rm{spot}}= 0.4-0.5$ 
to $r/R_{\rm{spot}}\simeq0.75$, $(\Delta B/\Delta \log \tau)_{\rm{0.0,-2.3}}$ is negative, with an average value of 
-- 40 (--100)\,G/$\log \tau$. Between $r/R_{\rm{spot}}\simeq 0.75$ 
and $r/R_{\rm{spot}}=1.0$ the $(\Delta B/\Delta \log \tau)_{\rm{0.0,-2.3}}$ is positive.

To crosscheck the presence of negative gradients in our results we performed inversions of the
\ionn{Si}{i}\,10827.1\,\AA\ and \ionn{Ca}{i}\,10833.4\,\AA\ lines separately, but this time the 
atmospheric model contained a height-independent magnetic field. Comparison of the retrieved magnetic 
field strength from the \ionn{Si}{i}\ line to that from  the \ionn{Ca}{i}\ line qualitatively 
confirms the vertical magnetic field gradient structure in the penumbrae obtained with the inversion
of the both lines together. For example: the magnetic field strength obtained from the inversion 
of the \ionn{Ca}{i}\,10833.4\,\AA\, line is 745\,G and that for the \ionn{Si}{i}\,10827.1\,\AA\ line is 730\,G. These values are for 
the location in the penumbra, $\rm{RF}_\mathnormal{B}$ of which are shown in Fig.~\ref{brf}(a)-(d). For the penumbral location
corresponding to $\rm{RF}_\mathnormal{B}$ displayed in Fig.~\ref{brf}(e)-(h) the retrieved magnetic field strength form the \ionn{Ca}{i}\,10833.4\,\AA\
and \ionn{Si}{i}\,10827.1\,\AA\ lines are 1190\,G and 1241\,G, respectively. The average formation heights for the \ionn{Ca}{i}\, and \ionn{Si}{i}\, lines
in the penumbral atmospheres are $\log \tau$ = --1.6 and $\log \tau$ = --2.1, respectively. These field strength values 
confirm the opposite signs of the gradients obtained by inverting both lines together.

\section{Hinode SOT/SP observations} \label{4_sot}

Low spatial resolution observations of a sunspot can lead to under- or overestimation of 
$B$ and $(\Delta B/\Delta \log \tau)$ due to unresolved multiple magnetic components with different
orientation of the magnetic vector within the resolution element. This effect is caused by signal 
cancellation of Stokes $V$ profiles. The unexpected result that 
$B$ decreases with optical depth, in azimuthal averages in the inner penumbra leads to the question if 
this outcome is a spurious result of the low spatial resolution of the VTT/TIP-2 data
or if it is real. 

To answer this question we analyze high spatial resolution Hinode SOT/SP 
observations. SOT/SP records full Stokes profiles of the \ionn{Fe}{i}\, 6301.5\,\AA\, and 
\ionn{Fe}{i}\, 6302.5\,\AA\, spectral lines. The SOT/SP observations  
analyzed here belong to active region NOAA 10933, which were taken on 
January 5, 2007 close to disk center (5\textdegree N, 2\textdegree W).  
This data set was inverted by \citet{Tiwari_2013,Tiwari_2015} and \citet{Vannoort_2013} 
and we employed the atmospheric parameters retrieved as part of these studies. These authors used the spatially coupled 
inversion technique of \citet{Vannoort_2012} to infer atmospheric 
parameters. The particular inversion considered here has been performed with an 
enhanced spatial sampling of $0\arcsecc 08$, compared to $0\arcsecc 16$ of the original Hinode 
SOT/SP \citep[see][for details]{Vannoort_2013}. The initial atmospheric model used by 
\citet{Tiwari_2013} consists of three nodes each for 
$T$, $v_{\rm{los}}$, $B$, $\gamma$, $\phi$ and $v_{\rm{mic}}$. 
The three nodes are at $\log \tau= 0.0, -0.9$ and $-2.5$.

The map of continuum intensity and $(\Delta B/\Delta \log \tau)$ 
maps are shown in Fig.~\ref{skt_maps}. The map of $(\Delta B/\Delta \log \tau)_{0.0,-0.9}$, (panel (b))
shows patches of negative values in the inner penumbra, i.e., at these locations $B$ 
decreases with optical depth. Patches of negative values also exist in the map of 
$(\Delta B/\Delta \log \tau)_{-0.9,-2.5}$, i.e.,  in the upper
atmospheric layer (see panels (c) of Fig.~\ref{skt_maps}), but they are 
smaller in number and with lower amplitude. For simplicity we shall call the atmosphere between $\log \tau= 0$ and --0.9 as the lower and 
between $\log \tau= -0.9$ and --2.5 as the upper layer. Just outside the visible boundary of 
the sunspot there is a ring of negative gradient in both the lower and upper atmospheric 
layers. This ring of negative gradient extends further in the upper layer compared 
to the lower layer.

The stratification of magnetic field strength with respect to $\log \tau$ at two different locations
in the penumbra are shown in Fig.~\ref{fe_rfb}. Panels (b) and (c) display $\rm{RF}_\mathnormal{B}$ of 
Stokes $I$ and $V$ profiles of the \ionn{Fe}{i}\, 6301.5\,\AA\, and \ionn{Fe}{i}\, 6302.5\,\AA\ lines for the atmosphere in panel (a), where
the magnetic field strength \textit{increases} with optical depth. Panels (e) and (f) represent those for 
the atmosphere depicted in panel (d) where the magnetic field strength \textit{decreases} with optical depth. $\rm{RF}_\mathnormal{B}$ for both the
atmospheres demonstrate that $\log \tau$ nodes used for the inversions of the \ionn{Fe}{i}\, 6301.5\,\AA\
and \ionn{Fe}{i}\, 6302.5\,\AA\ lines reasonably samples their formation heights.

\begin{figure}
\centering
   \includegraphics[width=0.48\textwidth]{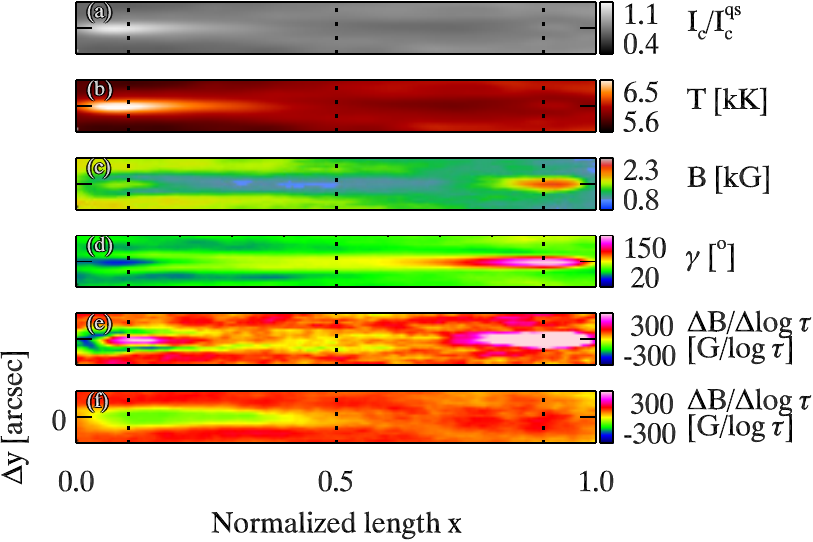}
      \caption{Physical properties of the standard filament from 
               \citet{Tiwari_2013}.
                  (a) Continuum intensity normalized to the quiet Sun continuum; 
                  (b) $T$ at $\log \tau= 0.0$;
                  (c) $B$ at $\log \tau= 0.0$;
                  (d) $\gamma$ at $\log \tau= 0.0$;
                  (e) $(\Delta B/\Delta \log \tau)_{0.0,-0.9}$ and
                  (f) $(\Delta B/\Delta \log \tau)_{-0.9,-2.5}$.
                   Left- and rightmost parts of the panels correspond 
                   to the inner and outer part of the filament, respectively,
                   with inner meaning closer to the umbra.
                   }
\label{skt_fil}
\end{figure}

\begin{figure}[!th]
\centering
   \includegraphics[width=0.48\textwidth]{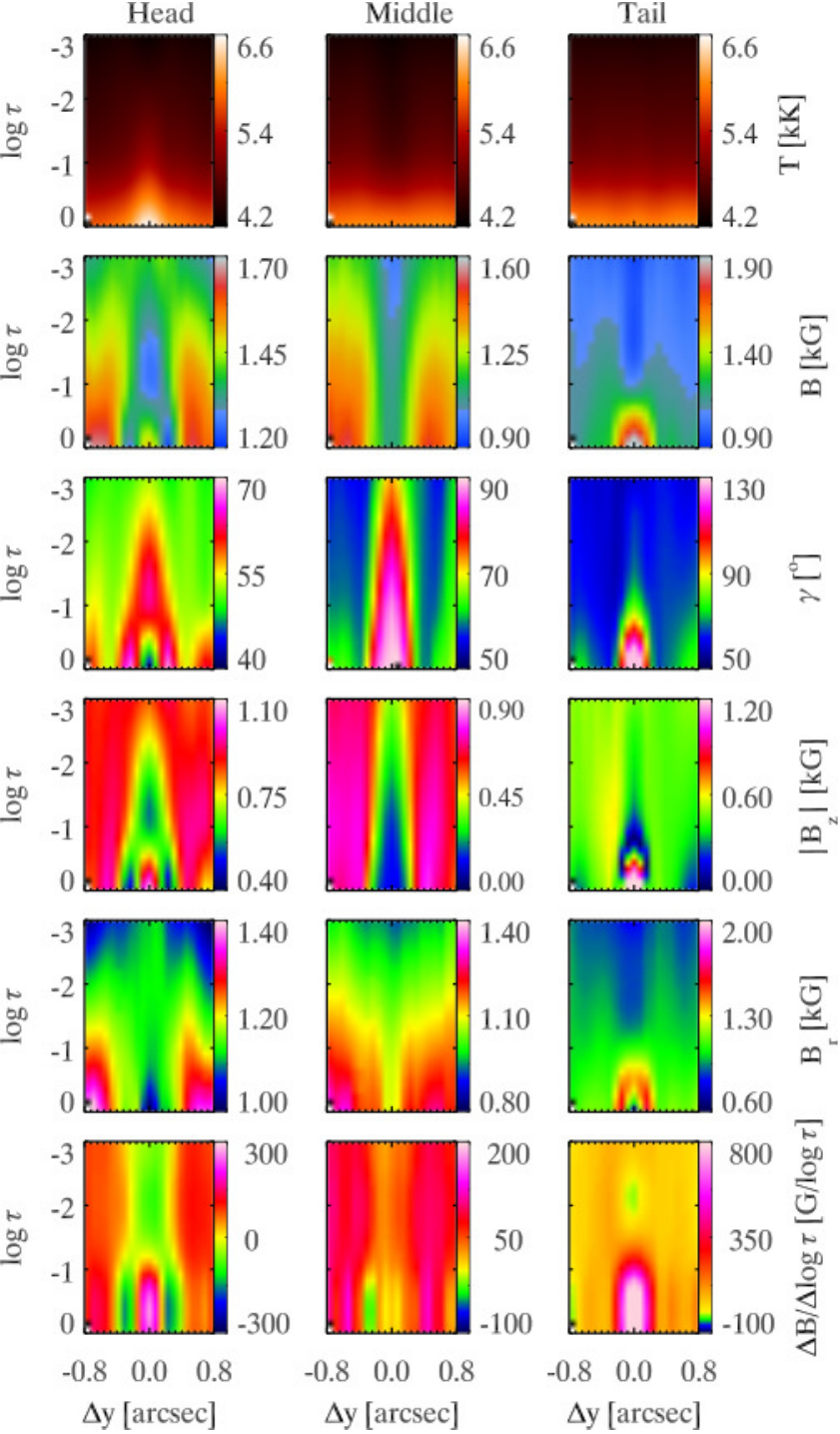}
      \caption{Optical depth stratification of various physical parameters 
               of the standard filament displayed in Fig.~\ref{skt_fil} 
               perpendicular to its axis. Panels in the first, second and 
               third column represent the head, middle and tail 
               of the filament respectively. From top to bottom (for each 
               column): $T$, $B$, $\gamma$, the vertical component of the
               magnetic field $|B_{\rm{z}}|$, the radial component of the magnetic 
               field $B_{r}$ and $(\Delta B/\Delta \log \tau)$. The head, middle 
               and tail positions are marked by \textit{dashed} vertical 
               lines in Fig.~\ref{skt_fil}.          
               }
\label{skt_fil_ver}
\end{figure}

\begin{figure}
\centering
   \includegraphics[width=0.48\textwidth]{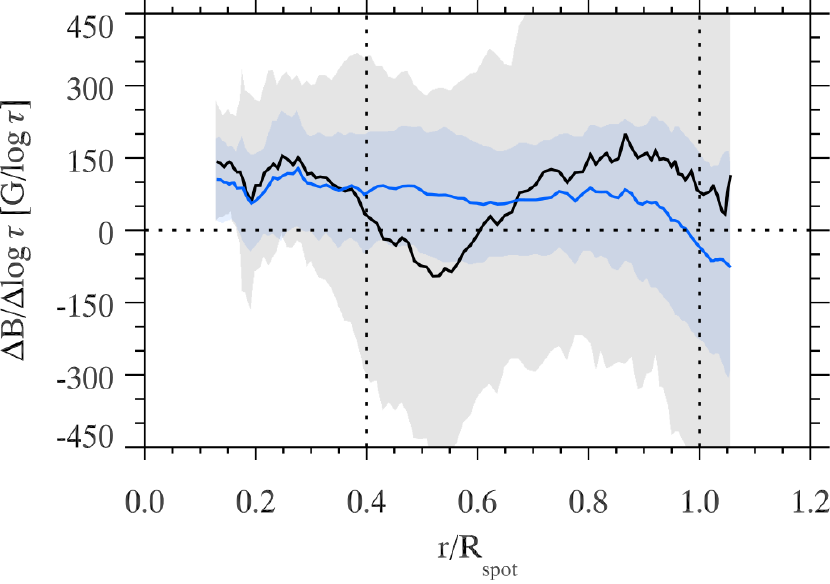}
      \caption{Azimuthally averaged magnetic field gradient as a function of normalized radial 
               distance, $r/R_{\rm{spot}}$, from the center of the sunspot observed with 
               SOT/SP. \textit{Black} and \textit{blue} curves depict 
               $(\Delta B/\Delta \log \tau)_{0.0,-0.9}$ and $(\Delta B/\Delta \log \tau)_{-0.9,-2.5}$,
               respectively.\,\textit{Shaded} areas represent the standard deviations.
               \,\textit{Dotted} vertical lines indicate
               the umbra-penumbra boundary and the outer boundary of the sunspot.
               }
\label{skt_radial} 
\end{figure}

\begin{figure}
\centering
\includegraphics[width=0.48\textwidth]{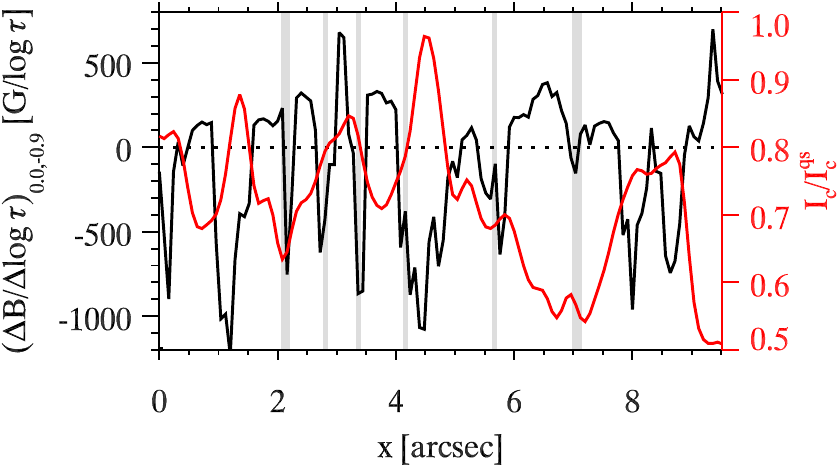}
      \caption{Fluctuations of  
               $(\Delta B/\Delta \log \tau)_{0.0,-0.9}$ (\textit{black} curve)
                and $I_{c}/I_{c}^{qs}$ (\textit{red} curve)
               along the \textit{white} line in Fig~\ref{skt_maps_roi}(a). \textit{Gray shaded} areas
               correspond to positions with polarity opposite to that in the 
               umbra. 
              }
\label{skt_slit}
\end{figure}

\begin{figure}
\centering
   \includegraphics[width=0.48\textwidth]{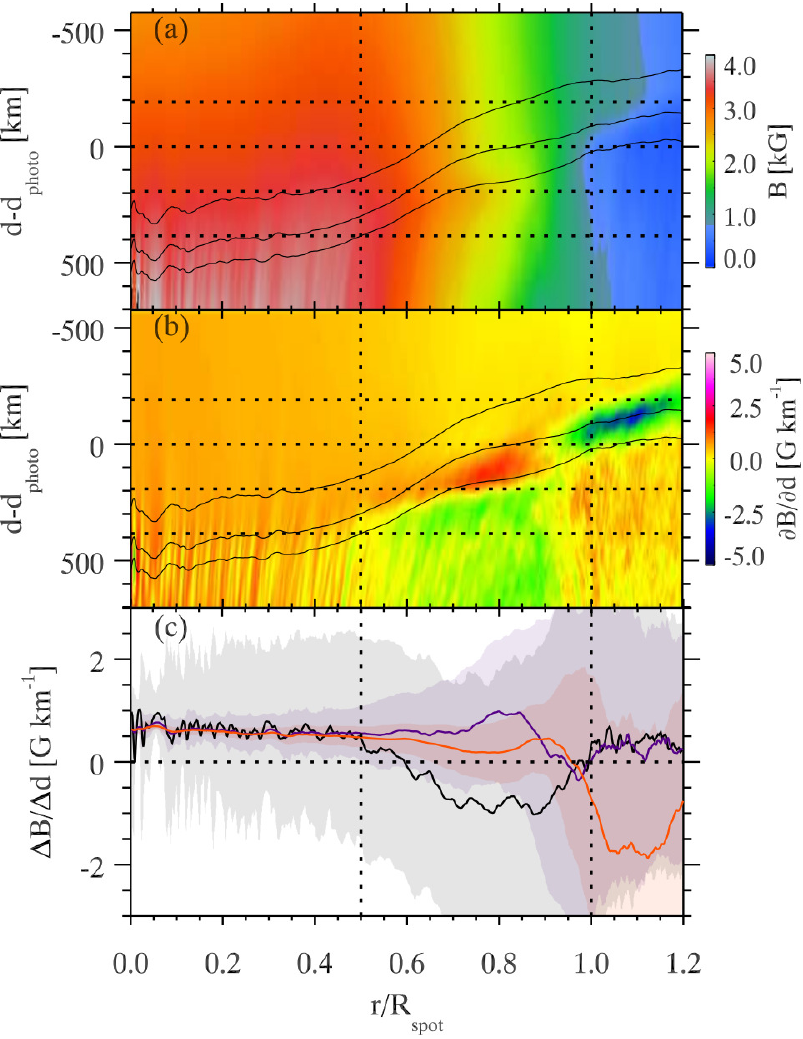}
      \caption{The azimuthally averaged magnetic field and its vertical gradient in the 3D MHD simulation. Panel (a) represents  
               the magnetic field  as a function of normalized radial 
               distance $r/R_{\rm{spot}}$ and geometrical depth $d-d_{\rm{photo}}$, here  $d_{\rm{photo}}$
               denotes the average geometrical depth of the quiet Sun photosphere at $\log \tau= 0$.
               Panel (b) displays $\partial B/\partial d$. \textit{Lower, middle} and \textit{upper}
               curves in panels (a) and (b) show the $\log \tau$ = 0.0, $\log \tau$ = --0.9 and $\log \tau$ = --2.5 levels,
               respectively.
               \textit{Black, blue} and \textit{red} curves in panel (c) depict 
               $(\Delta B/\Delta d)$ estimated between $d-d_{\rm{photo}}=384$\,km\, and 192\,km, 
               $d-d_{\rm{photo}}=192$\,km and 0\,km\, and $d-d_{\rm{photo}}= 0$\,km and --192\,km, respectively. These geometric 
               depth positions used in estimating $(\Delta B/\Delta d)$ are shown by horizontal lines in panels (a) and (b).  
               \,\textit{Shaded} areas represent standard deviations.\,\textit{Dotted} vertical lines in all panels indicate
               the umbra-penumbra boundary and the outer boundary of the sunspot.
               }
\label{OMHD_radial}
\end{figure}

The gradient maps are highly structured which can be seen 
in Fig.~\ref{skt_maps_roi}, a zoomed-in version of Fig.~\ref{skt_maps} with magnetic field inclination 
information added. 
Contours in panels (a) and (c)-(f) of Fig.~\ref{skt_maps_roi} separate patches 
with negative and positive values of $(\Delta B/\Delta \log \tau)_{0.0,-0.9}$. 
It is evident that the patches where the field has negative gradient mostly coincide 
with bright filaments. The gradient is steeper in the lower layer (panel (b)) 
compared to the upper layer (panel (c)). The gradient shows also fine structure 
within the body of individual filaments. Most parts of the filaments display negative
values of $(\Delta B/\Delta \log \tau)_{0.0,-0.9}$, while at the centers of filaments 
$(\Delta B/\Delta \log \tau)_{0.0,-0.9}$ is positive (indicated by an arrow in panel (b) of 
Fig.~\ref{skt_maps_roi}). In the upper layer the filaments have mostly negative 
$(\Delta B/\Delta \log \tau)_{-0.9,-2.5}$ along their central parts, in particular 
near their heads. Contours colored in \textit{white} in panel (b)
of Fig.~\ref{skt_maps_roi} show locations where the polarity of $B$ is opposite to that of 
the umbra at $\log \tau$ = 0.0. Narrow patches of opposite polarity at $\log \tau$ = 0.0 
are co-located with the patches of negative $(\Delta B/\Delta \log \tau)_{0.0,-0.9}$, 
but the former are narrower. Opposite polarity patches disappear
at $\log \tau$ = --2.5, whereas at $\log \tau$ = --0.9 opposite polarity patches
are located only at the tails of filaments.

Most of the filaments in our dataset display the properties mentioned above. 
To check weather these are generic, we consider the  
averaged penumbral filament constructed by \citet{Tiwari_2013}. They averaged 60 
penumbral filaments, 20 each from the inner, middle and outer penumbra. Prior to averaging they 
straightened  all filaments and then normalized their lengths. The inner, middle and outer 
filaments exhibit strong similarities in all physical properties, such as 
the plasma flow and the magnetic field. The temperature structure is also similar,
but does show some changes with the radial distance from the sunspot center. 
Some of the physical properties of this averaged filament are displayed in
Fig.~\ref{skt_fil}. The structure of the gradient of the averaged filament 
is similar to that noticed in individual filaments
(see Fig.~\ref{skt_maps_roi}), although much smoother. At the center, along the central ridge 
of the filament $(\Delta B/\Delta \log \tau)_{-0.0,-0.9}$ is positive,
surrounded by negative $(\Delta B/\Delta \log \tau)_{-0.0,-0.9}$ at the edges, in particular around the head.
The inner (i.e., near umbra) half ($x <0$.5 in Fig.~\ref{skt_fil}) 
of the filament has negative $(\Delta B/\Delta \log \tau)_{-0.9,-2.5}$ along its central ridge. The outer half ($x >0$.5) 
displays positive values of $(\Delta B/\Delta \log \tau)_{-0.9,-2.5}$. 

To figure out the penumbral magnetic field structure which can produce such 
a magnetic field gradient, it is important to see how the magnetic field is 
stratified with optical depth. The optical depth stratification of various physical 
parameters is shown in Fig.~\ref{skt_fil_ver}. As mentioned before, the atmosphere is fitted only on the three 
$\log \tau$ positions. The stratification at intermediate $\log \tau$ 
positions shown in Fig.~\ref{skt_fil_ver} is derived by interpolation using the same bicubic spline
approximation as used during the inversion procedure. The positions of 
these vertical cuts through the standard filament displayed in Fig.~\ref{skt_fil_ver} are indicated by 
\textit{dashed} vertical lines in Fig.~\ref{skt_fil}. 
In the central ridge ($\Delta y= 0$$\varcsec$) of the head of the filament (first column of Fig.~\ref{skt_fil_ver}), 
$B$ becomes weaker and more horizontal from $\log \tau = 0.0$ to $\log \tau= -0.9$ 
and $B$ increases again from $\log \tau= -0.9$ to $\log \tau= -2.5$, so that the vertical gradient of the field 
is positive between $\log \tau= 0.0$ and $\log \tau = -0.9$ and negative between $\log \tau = -0.9$ and
$\log \tau = -2.5$.
Between $\log \tau= 0.0$ and 
$\log \tau= -0.9$ the vertical component of the magnetic field, 
$B_{\rm{z}}$, also becomes weaker, whereas the radial component of the magnetic 
field, $B_{r}$, becomes stronger. 
At the edge  of the filament head ($\Delta y\simeq\pm 0\arcsecc4$), $B$ is horizontal and weaker in the deeper atmosphere 
and becomes more vertical and stronger in the upper atmosphere. The more vertical and
stronger field in the upper part of
the atmosphere is thought to be field from the surrounding spines, expanding and 
closing above the filament, which carries weaker magnetic fields. This might explain why we see a negative field gradient
at the location of the filament.

At the middle and tail part of the filament (second and third columns of 
Fig.~\ref{skt_fil_ver}) the horizontal field at the filament center is 
stronger compared to the overlying more vertical field. 
Hence in the middle and tail part of the bright filament 
the field gradient is positive.

\begin{figure*}
\centering
   \includegraphics[width=0.96\textwidth]{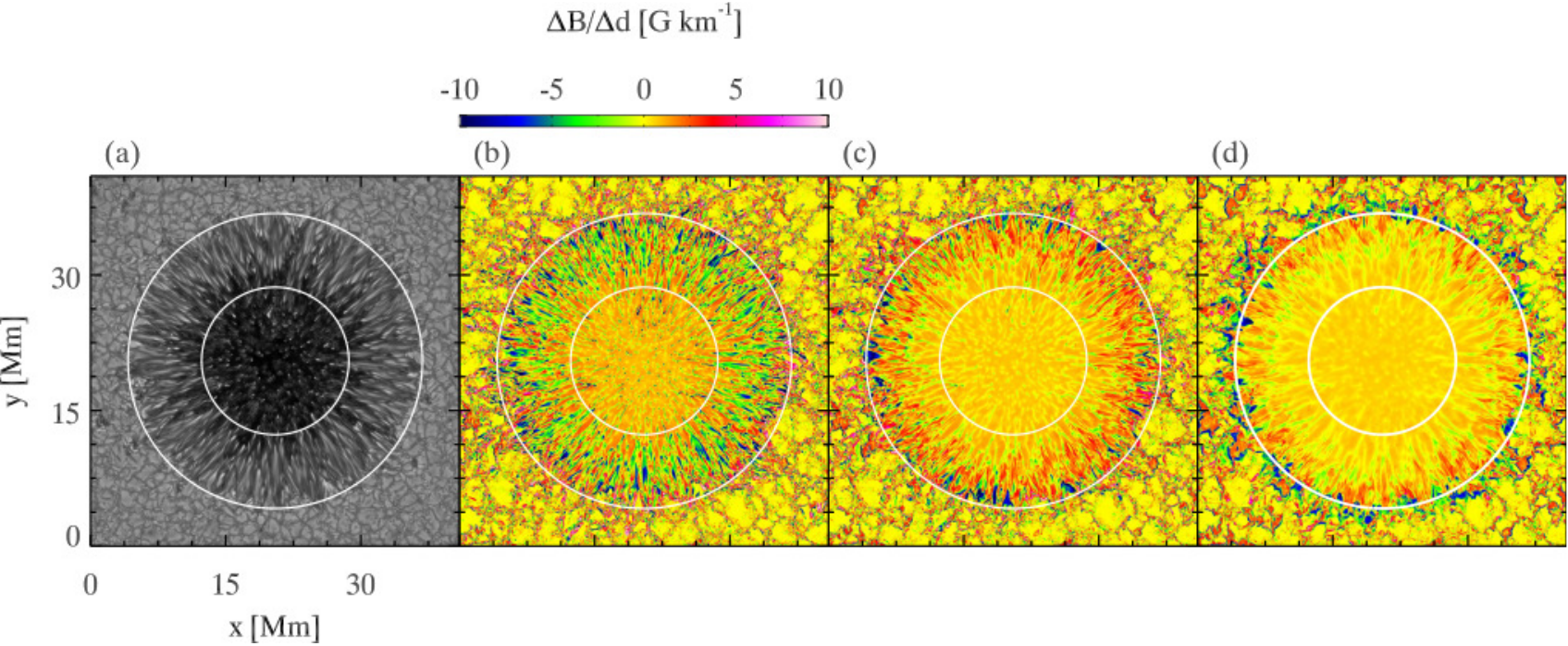}
      \caption{Continuum intensity (panel (a)) and the magnetic field gradient maps from
               a 3D MHD simulation of a sunspot computed by \citet{Rempel_2009a}. 
               Panels (b), (c) and (d) 
               depict $(\Delta B/\Delta d)$ estimated between
               $d-d_{\rm{photo}}=384$\,km\, and 
               192\,km, $d-d_{\rm{photo}}=192$\,km\, and 0 \,km\,
               and $d-d_{\rm{photo}}= 0$\,km\, and --192\,km, respectively.
               Here $d_{\rm{photo}}$ signifies the depth at which on average $\tau=0$
               is reached in the quiet Sun.
               \textit{White} circles in all panels represent the averaged umbra-penumbra 
               boundary and the outer boundary of the sunspot.
              }
\label{OMHD_maps}
\end{figure*}

\begin{figure*}
\centering
   \includegraphics[width=0.96\textwidth]{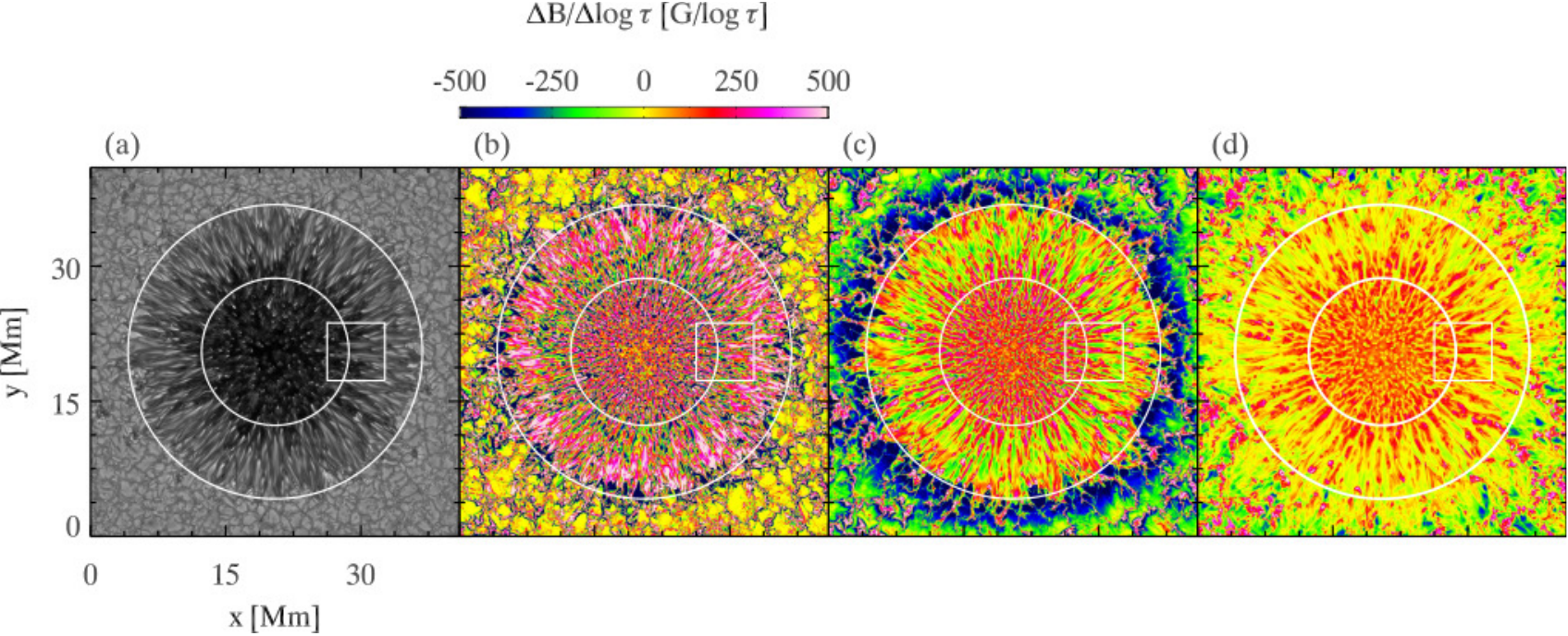}
      \caption{Continuum intensity (panel (a)) and $(\Delta B/\Delta \log \tau)$ maps obtained from
               the MHD simulation of a sunspot. Panels (b), (c) and (d) display 
               $(\Delta B/\Delta \log \tau)_{0.0,-0.9}$, $(\Delta B/\Delta \log \tau)_{-0.9,-2.5}$
               and $(\Delta B/\Delta \log \tau)_{-2.5,-3.5}$, respectively.               
               The \textit{white} circles have the same meaning as in Fig.~\ref{skt_maps}.
               The \textit{white} box denotes the location of the blowup displayed in Fig.~\ref{MHD_maps_roi}.
              }
\label{MHD_maps}
\end{figure*}

\begin{figure*}
\centering
   \includegraphics[width=0.96\textwidth]{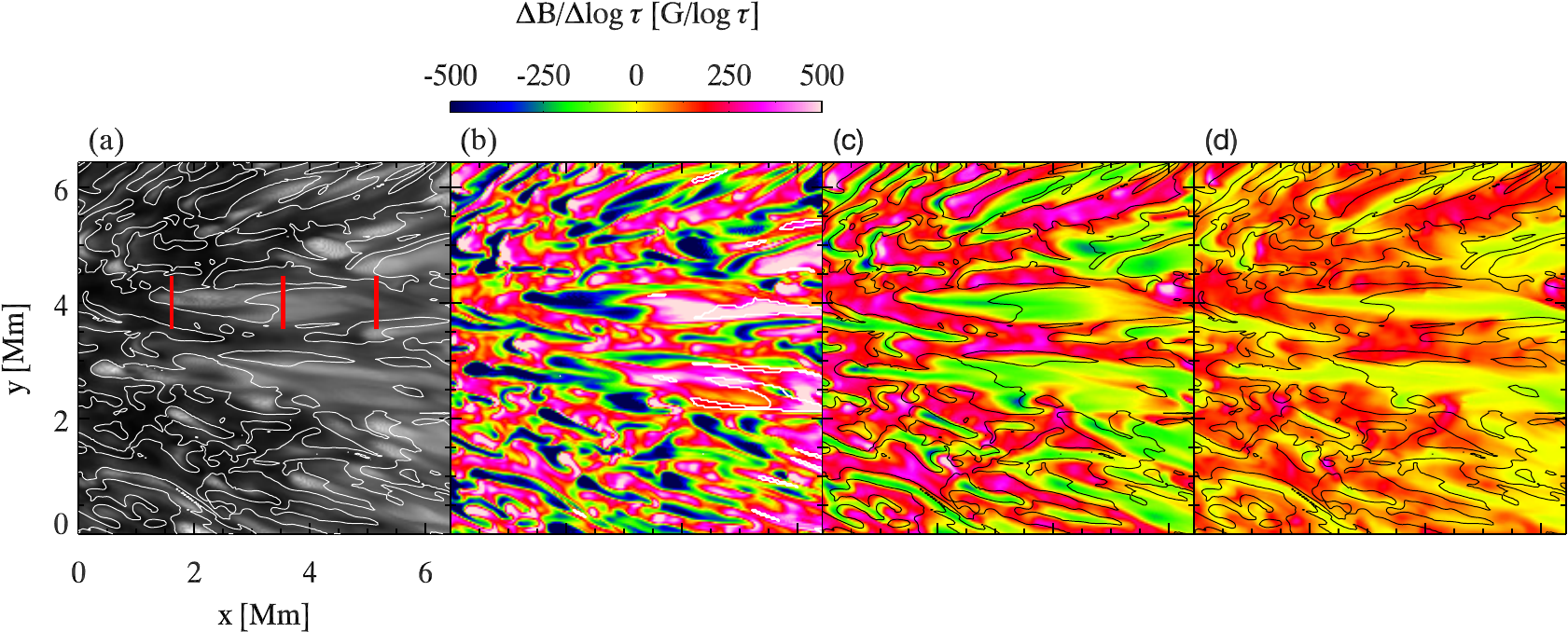}
      \caption{Blowups of the area within the
               \textit{white} boxes in Fig.~\ref{MHD_maps}.
               Contours in the panels (a), (c) and (d) encircles negative
               $(\Delta B/\Delta \log \tau)_{0.0,-0.9}$ and \textit{white} contours in
               the panel (b) indicate areas with polarity opposite to that in the 
               umbra. Three vertical bars in panel (a) mark the location of the vertical cuts
               displayed in Fig.~\ref{MHD_ver}. 
              }
\label{MHD_maps_roi}
\end{figure*}

To derive the dependence of $(\Delta B/\Delta \log \tau)$ on the radial distance 
from the sunspot center we use the same method as applied to the VTT/TIP-2 data 
(see Section~\ref{4_tip_res}). The radial dependence of $(\Delta B/\Delta \log \tau)_{0.0,-0.9}$ 
shows a qualitative similarity with that of $(\Delta B/\Delta \log \tau)_{0.0,-2.3}$ 
found from the VTT/TIP-2 observations, but with the gradients being more positive. Quantitatively, the positive values retrieved from the 
SOT/SP observations are higher: in the umbra the average value of $(\Delta B/\Delta \log \tau)_{0.0,-0.9}$  
is found to be 150\,G/$\log \tau$ (see Fig.~\ref{skt_radial}), 
$\sim 1.4$\,G\,km$^{-1}$ in the geometrical depth scale. In the inner penumbra, 
between $r\simeq 0.43 R_{\rm{spot}}$ and $r\simeq 0.63 R_{\rm{spot}}$, $(\Delta B/\Delta \log \tau)_{0.0,-0.9}$ 
is negative, with the maximum negative value of --80\,G/$\log \tau$ being reached at $r= 0.5 R_{\rm{spot}}$. 
$(\Delta B/\Delta \log \tau)_{0.0,-0.9}$ is positive beyond $r\simeq 0.6 R_{\rm{spot}}$. 
$(\Delta B/\Delta \log \tau)_{-0.9,-2.5}$ amounts to 120\,G/$\log \tau$ 
on average in the umbra and remains positive also in the penumbra. Outside the boundary of the
sunspot, on average $(\Delta B/\Delta \log \tau)_{-0.9,-2.5}$ is negative. 

Fig.~\ref{skt_slit} displays $(\Delta B/\Delta \log \tau)_{0.0,-0.9}$ and $I_{\rm c}/I_{\rm c}^{\rm qs}$
along a slit (corresponding to the horizontal white line in Fig.\ref{skt_maps_roi}(a)) 
in the penumbra cutting across several penumbral filaments. $I_{\rm c}$ and  $I_{\rm c}^{\rm qs}$ represent continuum intensity and 
averaged continuum intensity in the quiet Sun, respectively.
Along the slit both positive and negative patches of $(\Delta B/\Delta \log \tau)_{0.0,-0.9}$ are 
present. At the location of filaments $(\Delta B/\Delta \log \tau)_{0.0,-0.9}$ has negative values with
larger magnitudes than the positive $(\Delta B/\Delta \log \tau)_{0.0,-0.9}$ of the dark background.  
Hence, the azimuthal average of $(\Delta B/\Delta \log \tau)_{0.0,-0.9}$ turns out to be negative in the inner
penumbra.

\section{Comparison with 3D MHD simulations} \label{4_mhd}

To better understand the origin of the azimuthally averaged
negative $(\Delta B/\Delta \log \tau)_{0.0,-0.9}$ and $(\Delta B/\Delta \log \tau)_{0.0,-2.3}$ in sunspot penumbrae 
detected in the SOT/SP and VTT/TIP-2 observations, respectively, we analyzed a snapshot from a 3D
MHD simulation of a sunspot by \citet{Rempel_2009a}; cf. \citet{Rempel_2011}. 
The advantage of studying the MHD simulation is that we can determine the gradient also on a
geometrical depth scale, $d$, rather than just on an optical depth scale. 

The azimuthally averaged $B$ and $\partial B/\partial d$ stratified with respect to 
geometrical depth are depicted in Fig.~\ref{OMHD_radial}(a) and (b), respectively.
In Fig.~\ref{OMHD_radial}, $d$ denotes the geometrical depth ($d$ increases into the Sun)
and $d_{\rm{photo}}$ is the average geometrical 
depth in the quiet Sun photosphere at $\log \tau= 0$.
The considerable difference between the geometrical depth scale (shown by \textit{dashed} lines) 
to the optical depth scale (shown by \textit{solid} lines) is clearly evident.
$\partial B/\partial d$ is positive above $\log \tau= 0$ for $r/R_{\rm{spot}} < 0.9$ 
and negative for  $r/R_{\rm{spot}} > 0.9$. This is in contradiction with the azimuthally averaged 
radial profile of $(\Delta B/\Delta \log \tau)_{0.0,-0.9}$ obtained from SOT/SP observations, which have negative values in the inner penumbra.
In the MHD simulation the azimuthally averaged $\partial B/\partial d$ is negative in the penumbra only below $\log \tau= 0$.  
With the MHD simulation we have the advantage that we can analyze the complete stratification of $\partial B/\partial d$,
whereas from the observation magnetic field values are known only as a function of optical depth (and at discrete nodes). Any conversion to geometrical height implies additional assumptions. We also estimate    
$(\Delta B/\Delta d)$, which denotes the vertical gradient of the magnetic field between given geometrical 
depths from the MHD simulation (see $dashed$ horizontal lines in Fig.~\ref{OMHD_radial}(a) and (b)). $(\Delta B/\Delta d)$ is defined as follows,

\begin{equation} \label{eqn_5.3}
{\left(\frac{\Delta B}{\Delta d}\right)_{\rm{p,q}}=\frac{B(p)-B(q)}{p-q}} \,,
\end{equation}

\noindent
where $p$ and $q$ denotes the lower and upper geometrical depth, $d-d_{\rm{photo}}$, respectively. 

Azimuthal averages of $(\Delta B/\Delta d)$, calculated for three depth intervals are
presented in Fig.~\ref{OMHD_radial}(c). The $\log \tau= 0$ surface in the inner penumbra 
in the MHD simulation is on average $\sim400$\,km deeper than in the quiet Sun, so we do not 
look at gradients of $B$ below $d-d_{\rm{photo}}=384$\,km.
It shows an absence of negative $(\Delta B/\Delta d)_{192,0}$ and $(\Delta B/\Delta d)_{0,-192}$
within the sunspot although these gradients achieve negative values outside the boundary of the sunspot.
The gradient in the lowest considered layer, $(\Delta B/\Delta d)_{384,192}$, exhibits
negative values for $0.9\gtrsim r/R_{\rm{spot}}\gtrsim0.6$. 
The shaded areas in Fig.~\ref{OMHD_radial}(c) show standard deviations of $(\Delta B/\Delta d)$ 
along the circles used to estimate the azimuthal averages.

\begin{figure}
\centering
   \includegraphics[width=0.48\textwidth]{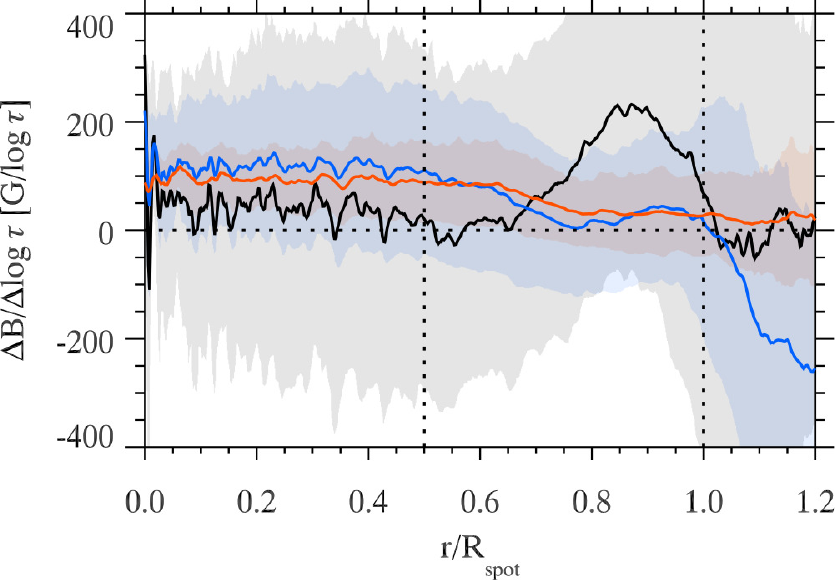}
      \caption{Azimuthally averaged $(\Delta B/\Delta \log \tau)$ as a function of normalized radial 
               distance $r/R_{\rm{spot}}$, from the MHD simulation of a sunspot.
               \textit{Black, blue and red} curves depict 
               $(\Delta B/\Delta \log \tau)_{0.0,-0.9}$, $(\Delta B/\Delta \log \tau)_{-0.9,-2.5}$
               and $(\Delta B/\Delta \log \tau)_{-2.5,-3.5}$, respectively.
               \,\textit{Shaded} areas represent standard deviations.\,\textit{Dotted} 
               vertical lines indicate the
               umbra-penumbra boundary and the outer boundary of the sunspot.
               }
\label{MHD_radial}
\end{figure}

Fig.~\ref{OMHD_maps} shows a continuum intensity map at 6303.1 \AA\, (panel (a)) and maps of $(\Delta B/\Delta d)_{384,192}$, 
$(\Delta B/\Delta d)_{192,0}$ and $(\Delta B/\Delta d)_{0,-192}$ in panels (a)-(d), respectively. There are patches where 
$(\Delta B/\Delta d)_{384,192}$ is negative for $0.9\gtrsim r/R_{\rm{spot}}\gtrsim0.6$. 
We see fewer of patches in the penumbra where $(\Delta B/\Delta d)_{192,0}$ or $(\Delta B/\Delta d)_{0,-192}$ is negative 
compared to $(\Delta B/\Delta \log \tau)_{0.0,-0.9}$ in the Hinode SOT/SP and VTT/TIP-2 observations.

As mentioned above, the azimuthally averaged $(\Delta B/\Delta d)_{192,0}$ and $(\Delta B/\Delta d)_{0,-192}$ 
in the MHD simulation does not display negative gradients in the inner penumbra whereas the SOT/SP and 
VTT/TIP-2 observations show that the vertical gradient (in optical depth scales) of the 
inner penumbra is negative in the azimuthal averages. 
There are multiple possible reasons for this discrepancy between the MHD simulation 
and the observations:

\begin{enumerate}
 \item The gradients of $B$ related to geometrical depth may not be directly comparable to 
       gradients at a given optical depth (since iso-optical depth surfaces are likely strongly 
       corrugated in the azimuthal direction, depending on the field strength and the temperature).
       Also, a given optical depth samples different geometrical depths at various radii in the penumbra.
 \item There are shortcomings in the simulations, which lead to a departure from realism 
       in the synthetic sunspot's magnetic structure. For example, the penumbra is rather shallow in 
       the MHD simulation and the lowest layer analyzed here ($d-d_{\rm{photo}}=384$\,km\,) is below the lower boundary 
       of the magnetic field at some locations at $r/R_{\rm{spot}}\gtrsim0.6$. This may explain 
       some of the negative $(\Delta B/\Delta d)_{384,192}$ at larger  $r/R_{\rm{spot}}$ in the simulation.
 \item The inversions of the observed Stokes profiles may be imperfect regarding the 
       magnetic structure, although the inversions of the two rather diverse observed data 
       sets (different spectral lines, different spatial resolution, ground-based vs.
       space-borne, etc.) analyzed using contrasting approaches (straylight and PSF removed vs.
       non consideration of straylight) result in qualitatively similar magnetic field gradients.  
 
\end{enumerate}

For a comparison of the MHD atmosphere with the VTT/TIP-2 and SOT/SP observations according to point (1),
we computed $(\Delta B/\Delta \log \tau)$ between two iso-$\tau$ surfaces, determined using the SPINOR code,
everywhere in the simulated sunspot following Eqn.~\ref{eqn_5.2}.
The resulting maps are shown in Fig.~\ref{MHD_maps}. The $(\Delta B/\Delta \log \tau)_{0.0,-0.9}$ 
map reveals patches with negative values between $r/R_{\rm{spot}}\approx0.4$ and 0.7, similar to what is seen in 
the SOT/SP observations. Patches with negative values in the $(\Delta B/\Delta \log \tau)_{-0.9,-2.5}$ 
and $(\Delta B/\Delta \log \tau)_{-2.5,-3.5}$ maps are, however, shifted to larger radii, lying between $r/R_{\rm{spot}}\approx0.7$ 
and $r/R_{\rm{spot}}\approx1.0$. This behavior is not visible in the SOT/SP data. 
There the patches with negative values of $(\Delta B/\Delta \log \tau)$ are mostly found in 
the inner penumbra, also for the upper layer (see Fig.~\ref{skt_maps}). 
At the boundary of the sunspot we see a ring where 
$(\Delta B/\Delta \log \tau)_{0.0,-0.9}$ is negative. A much more prominent such ring is located further out in 
$(\Delta B/\Delta \log \tau)_{-0.9,-2.5}$ map. This is likely associated with the magnetic canopy surrounding the sunspot.

\begin{figure}
\centering
   \includegraphics[width=0.48\textwidth]{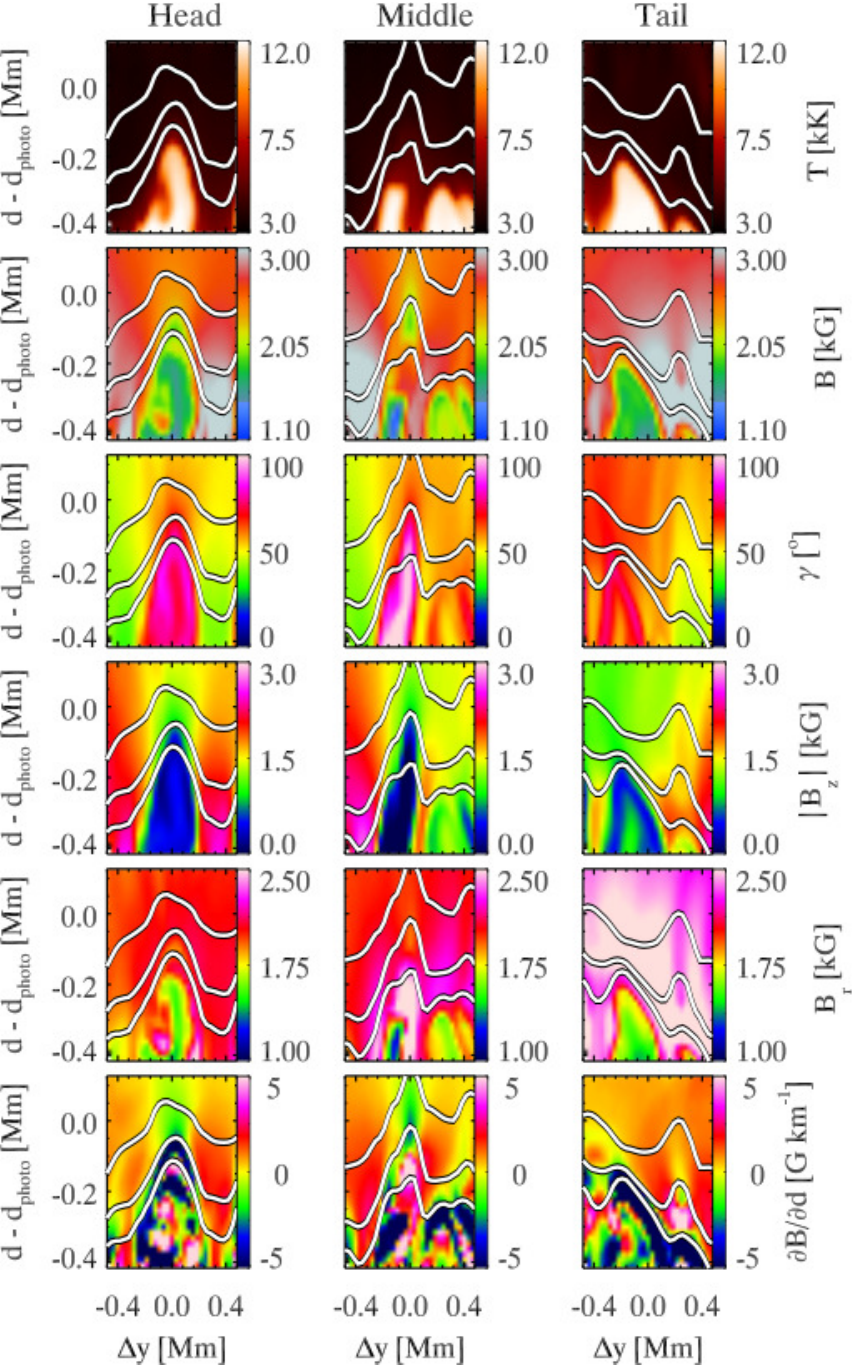}
      \caption{Geometrical depth stratification of various physical parameters 
               perpendicular to the axis of a filament in the MHD simulation.
               The locations of the plotted cuts are indicated by $red$ bars in 
               Fig.~\ref{MHD_maps_roi}(a). Panels in the first, second and 
               third column represent the head, middle and tail 
               of the filament, respectively. From top to bottom (for each 
               column): $T$, $B$, $\gamma$, 
               $\lvert B_{\rm{z}}\rvert$, $B_{\rm{r}}$  and $\partial B/\partial d$ are plotted.
               \textit{Lower, middle} and \textit{upper white} curves in all panels 
               show $\log \tau =  0.0, -0.9, -2.5$ levels, respectively.
              }
\label{MHD_ver}
\end{figure}

\begin{figure}
\centering
   \includegraphics[width=0.48\textwidth]{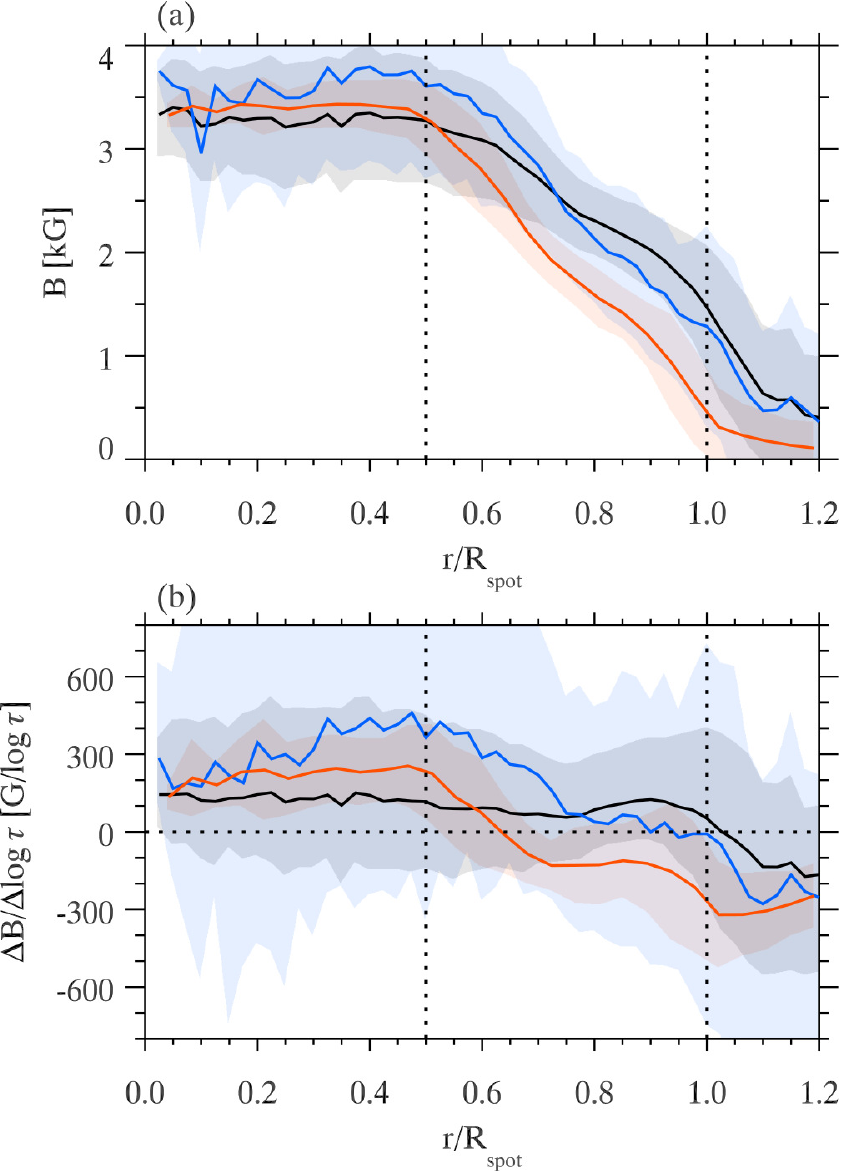}
      \caption{Radial dependence of $B$ at 
                $\log \tau= 0$ (panel (a)) and $(\Delta B/\Delta \log \tau)_{0.0,-2.3}$ (panel(b))
                from the MHD simulation.
                \textit{Black} curves are estimated directly from the MHD simulation, 
                \textit{blue} curves are results of inversions of synthesized 
                Stokes profiles from the MHD simulation at native resolution. 
                \textit{Red} curves are from inversions of synthesized 
                Stokes profiles from the MHD simulation degraded to a spatial 
                resolution of $\sim1\arcsecc0$.\,\textit{Shaded} areas
                represent standard deviation. The \textit{dotted} 
                vertical lines in both panels indicate the 
                umbra-penumbra boundary and the outer boundary of the sunspot.
                }
\label{MHD_radial_comp}
\end{figure}

Blowups of the region inside the white squares in Fig.~\ref{MHD_maps} are displayed in Fig.~\ref{MHD_maps_roi}.
In the MHD simulation too, patches where  $(\Delta B/\Delta \log \tau)_{0.0,-0.9}$ is negative, coincide with bright 
filaments. In the lower layers, patches with negative $(\Delta B/\Delta \log \tau)$ exist only around the heads
of the filaments. In the upper layers large parts of the filament's body 
show a weak negative $(\Delta B/\Delta \log \tau)$.

The azimuthally averaged $(\Delta B/\Delta \log \tau)$ values as derived from the MHD simulation, 
again as a function of the normalized sunspot radius, are shown in Fig.~\ref{MHD_radial}. The 
shaded areas represent standard deviations. The radial dependence of $(\Delta B/\Delta \log \tau)_{0.0,-0.9}$ 
is qualitatively similar to that of the SOT/SP observations showing negative values (--25\,G/$\log \tau$) in 
the inner penumbra. However these negative gradients are much weaker than in the observations. 
In the middle and outer penumbra $(\Delta B/\Delta \log \tau)_{0.0,-0.9}$ is positive but with  
some quantitative differences to the SOT/SP observations. The azimuthally averaged $(\Delta B/\Delta \log \tau)_{-0.9,-2.5}$ 
in the MHD simulation is positive
at all radial distances within the sunspot, which is similar to that of the SOT/SP observations.
Outside the sunspot $(\Delta B/\Delta \log \tau)_{-0.9,-2.5}$ is negative.
The standard deviations of the fluctuations of \textit{B} in the 
azimuthal averages have higher values in the MHD simulation compared 
to the SOT/SP observations. We attribute this to the higher spatial 
resolution of the MHD data. The azimuthal averages of the vertical gradient of the magnetic field obtained 
from VTT/TIP-2, SOT/SP and the MHD simulation is compared in Table~\ref{tab5.2}. 
The values listed in the table are averaged over the respective areas.

The azimuthally averaged $(\Delta B/\Delta \log \tau)_{0.0,-0.9}$ in the MHD simulation 
shows some negative values in the inner penumbra, the unlike azimuthally averaged $\partial B/\partial d$ 
which has only positive values above $\log \tau = 0$ in the inner penumbra (see Fig.~\ref{OMHD_radial}(b)).
Also, the spatially resolved maps of the vertical gradient
of the magnetic field have negative values patches in the geometrical as well as in optical 
depth scale, which predominantly coincide with the inner part (head) of the penumbral filaments. 
This suggests that the corrugation of the optical depth surfaces may play some role in producing the negative $(\Delta B/\Delta \log \tau)_{0.0,-0.9}$ 
in the azimuthal averages in the inner penumbra.

To investigate the effect of the corrugation of the optical depth surfaces on the azimuthally averaged
$(\Delta B/\Delta \log \tau)_{0.0,-0.9}$,  the height stratifications of the physical properties at the head, middle and 
tail of a penumbral filament from the MHD simulation are plotted in Fig.~\ref{MHD_ver}.
At the head of the filament $\partial B/\partial d$ has negative values between $\log \tau = 0.0$ (lower white curve)
and $\log \tau = -0.9$ (middle white curve), whereas it is positive in the surroundings (spines) of the filaments. 
The negative $\partial B/\partial d$ at the head of the filaments is caused by the closing of the stronger vertical fields
from the spines above the weaker and relatively horizontal fields of the filament (inter-spines).
The negative $\partial B/\partial d$ above $\log \tau = 0.0$ is more prominent at the head of the filament
compared to the middle and tail of the filament. At the head of the filament, the extension of strong negative $\partial B/\partial d$ with 
geometrical depth follows the corrugation of $\log \tau = 0.0$ and $\log \tau = -0.9$ levels. So, in this case areas with 
negative gradients (filaments) contribute more to the azimuthal averages of the vertical gradient of the magnetic field 
(in the geometrical depth scale) than areas with positive gradients (spines). Since there are more filament heads in the inner penumbra 
than in the middle and outer penumbra, the azimuthally averaged $(\Delta B/\Delta \log \tau)_{0.0,-0.9}$ is negative 
in the inner penumbra, as found in the SOT/SP and VTT/TIP-2 observations as well as in the MHD simulation.\\

\begin{table*}

\caption{Comparison of spatially averaged $(\Delta B/\Delta \log \tau)$ values obtain form VTT/TIP-2, SOT/SP and the MHD simulation in terms
        of the inner- and outer halves of the penumbra.\label{tab5.2}}
\begin{center} 
\begin{tabular}{c c c c c}
\hline\hline
   & VTT/TIP-2 & SOT/SP & MHD  \\
\hline
Inner &  $\left(\frac{\Delta B}{\Delta \log \tau}\right)_{\rm{0.0,-2.3}}\simeq-70\,\rm{G/\log \tau}$ &
                   \begin{tabular}{c} $\left(\frac{\Delta B}{\Delta \log \tau}\right)_{\rm{0.0,-0.9}}\simeq-20\,\rm{G/\log \tau}$ \\ $\left(\frac{\Delta B}{\Delta \log \tau}\right)_{\rm{-0.9,-2.5}}\simeq73\,\rm{G/\log \tau}$\end{tabular} &
                   \begin{tabular}{c} $\left(\frac{\Delta B}{\Delta \log \tau}\right)_{\rm{0.0,-0.9}}\simeq5\,\rm{G/\log \tau}$ \\ $\left(\frac{\Delta B}{\Delta \log \tau}\right)_{\rm{-0.9,-2.5}}\simeq65\,\rm{G/\log \tau}$\end{tabular} \\
\hline
Outer &  $\left(\frac{\Delta B}{\Delta \log \tau}\right)_{\rm{0.0,-2.3}}\simeq9\,\rm{G/\log \tau}$ &
                   \begin{tabular}{c} $\left(\frac{\Delta B}{\Delta \log \tau}\right)_{\rm{0.0,-0.9}}\simeq137\,\rm{G/\log \tau}$ \\ $\left(\frac{\Delta B}{\Delta \log \tau}\right)_{\rm{-0.9,-2.5}}\simeq52\,\rm{G/\log \tau}$\end{tabular} &
                   \begin{tabular}{c} $\left(\frac{\Delta B}{\Delta \log \tau}\right)_{\rm{0.0,-0.9}}\simeq172\,\rm{G/\log \tau}$  \\ $\left(\frac{\Delta B}{\Delta \log \tau}\right)_{\rm{-0.9,-2.5}}\simeq24\,\rm{G/\log \tau}$\end{tabular} \\

\hline
\end{tabular}
\end{center}
\end{table*}

\subsection{The effect of spatial resolution on $(\Delta B/\Delta \log \tau)$}

To understand the effect of the simplified model atmosphere used to fit the
Stokes profiles observed with VTT/TIP-2, we inverted Stokes profiles 
of the \ionn{Si}{i}\,10827\,\AA\, and \ionn{Ca}{i}\,10833\,\AA\, spectral lines
synthesized along vertical rays passing through every ($x,y$) grid point of the
MHD simulation. The inversions start from the same set of initial parameters as 
used to fit the VTT/TIP-2 data. One set of inversions is carried out at the native 
resolution of the MHD simulation, another after degrading the synthesized Stokes 
images by a point spread function (PSF) which has a Gaussian shape with FWHM of 
$1.\arcsec0$, close to the spatial resolution of the VTT/TIP-2 data presented in Section~\ref{4_tip}.
In the latter case, Stokes images are synthesized after binning the MHD atmosphere such that
the resultant pixel size is equal to that in the VTT/TIP-2 data.

In Fig.~\ref{MHD_radial_comp}, the radial dependence of $B$ (at $\log \tau= 0$)
and $(\Delta B/\Delta \log \tau)_{0.0,-2.3}$, deduced from the inversions of the synthetic data
are compared with the original $B$ and $(\Delta B/\Delta \log \tau)_{0.0,-2.3}$ values obtained from 
the MHD simulation. From the sunspot center to half of the sunspot radius, $B$ is reproduced 
reasonably by the inversions of synthesized Stokes profiles, both at native and degraded 
spatial resolution of the MHD simulation. The inversions tend to overestimate 
both $B$ and its vertical gradient. In the outer half of the sunspot $B$ 
is underestimated by $\sim200$\,G when we perform inversions of Stokes profiles synthesized in 
native spatial resolution. The inversion of spatially degraded Stokes profiles results in a 
further underestimate of $B$ by up to $\sim1000$\,G, mainly as a result of signal cancellation in
unresolved structures. Between $r/R_{\rm{spot}}\simeq 0.20$ and $r/R_{\rm{spot}}\simeq 0.70$ the inversions 
with the simplified model at native resolution overestimate $(\Delta B/\Delta \log \tau)_{0.0,-2.3}$, while between 
$r/R_{\rm{spot}}\simeq 0.75$ and $r/R_{\rm{spot}}=1.0$ they underestimate 
$(\Delta B/\Delta \log \tau)_{0.0,-2.3}$, but displays the correct sign.
The inversion results of the spatially degraded Stokes profiles show a qualitative difference to the true 
gradient in the model, leading to negative $(\Delta B/\Delta \log \tau)_{0.0,-2.3}$ in the outer penumbra. 
The main result, from the VTT/TIP-2 data, namely the dominant presence of
negative $(\Delta B/\Delta \log \tau)_{0.0,-2.3}$ in parts of the penumbra ,
is reproduced, although, the negative values of 
$(\Delta B/\Delta \log \tau)_{0.0,-2.3}$ occur in the outer part of the penumbra in 
the MHD simulation, while they are located in the inner penumbra in the observations. 

This experiment demonstrates that a lower spatial resolution can lead to apparent negative  
$(\Delta B/\Delta \log \tau)$ in the penumbra. One possibility is that unresolved opposite polarity
patches in the penumbra could significantly influence $(\Delta B/\Delta \log \tau)$ deduced from observations,
especially, if the population of opposite polarity patches changes with height in the photosphere.

\begin{figure}
\centering
   \includegraphics[width=0.48\textwidth]{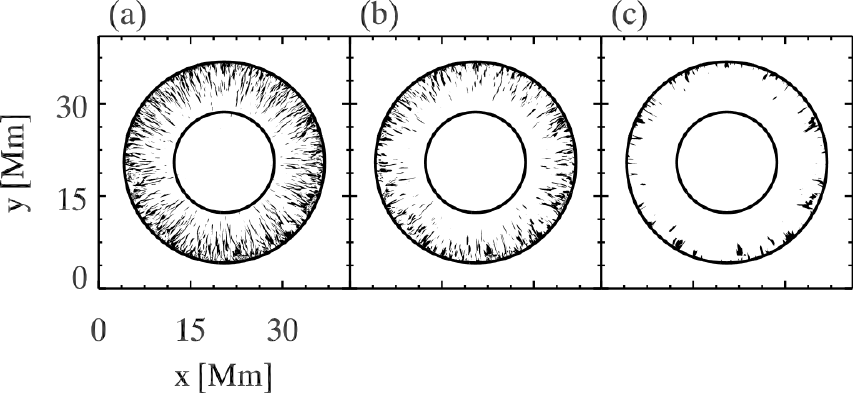}
      \caption{Maps of opposite polarity patches from the MHD
              simulation of a sunspot.
              ~\textit{Black} areas indicate positions of opposite
              magnetic polarity. Panels (a), (b) and (c) correspond to
	      $\log \tau = 0$, $\log \tau = -0.9$ and
	      $\log \tau= -2.5$, respectively. 
	      Circles in all panels represent the umbra-penumbra boundary and
	      the outer boundary of the sunspot.
              }
\label{mhd_pol}
\end{figure}

\begin{figure}
\centering
   \includegraphics[width=0.48\textwidth]{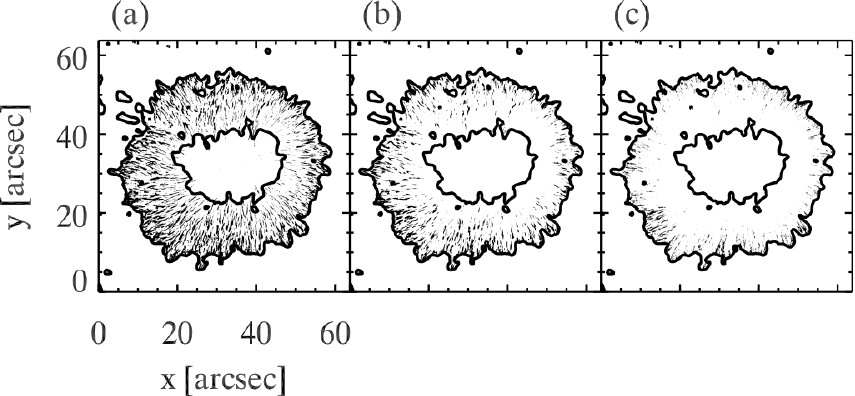}
      \caption{Same as Fig.~\ref{mhd_pol} but for SOT/SP observations.
              } 
\label{skt_pol}
 \end{figure}

\begin{figure}
\centering
   \includegraphics[width=0.48\textwidth]{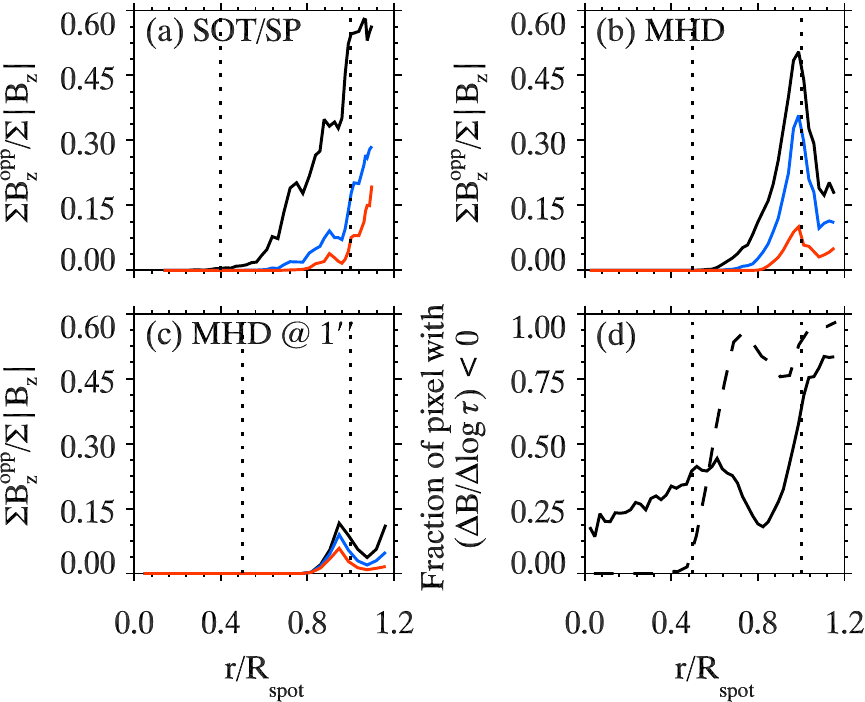}
      \caption{Panel (a) shows the radial distribution of opposite polarity flux 
               $B_{\rm{z}}^{\rm{opp}}$ as a fraction of the net flux \textbar$B_{\rm{z}}$\textbar\, 
               from the SOT/SP observations. Panel (b) displays 
               the radial distribution of opposite polarity flux for the MHD simulation. Panel (c) 
               corresponds to opposite flux from inversions of synthesized Stokes
               profiles from the MHD simulation which was degraded with a Gaussian 
               of $1.\arcsec0$ width. \textit{Black, blue} and \textit{red} curves correspond to 
               $\log \tau= 0.0$, $\log \tau= -0.9$ and $\log \tau= -2.5$, respectively.
               \textit{Solid} and \textit{dashed} curves in panel (d) display the numbers of pixels where
               $(\Delta B/\Delta \log \tau)_{0.0,-2.3} < 0$ at native and degraded spatial resolution in the MHD
               simulation, respectively.\,\textit{Dotted} vertical lines in all the panels indicate the 
               umbra-penumbra boundary and the outer boundary of the sunspot.
               }
\label{opp_flux}
\end{figure}


\section{Opposite polarity patches in penumbrae} \label{4_polarity}

In this section we analyze the distribution of opposite polarity\- patches at different $\log \tau$
surfaces, both in the MHD simulation and the SOT/SP observations, in order to understand the relation 
between opposite polarity patches and the presence of negative $(\Delta B/\Delta \log \tau)$

The positions of patches with polarity opposite to that in the umbra at three $\log \tau$ 
levels (0.0,--0.9 and --2.5) in the MHD simulation are shown in Fig.~\ref{mhd_pol}. 
At $\log \tau= 0$ the opposite polarity patches appear at almost all radial distances 
in the penumbra with the number of patches increasing towards the outer penumbra. 
At $\log \tau= -0.9$ and $\log \tau= -2.5$ the opposite polarity 
patches are mainly restricted to the middle to outer penumbra and to the outer penumbra, 
respectively. The positions of the opposite polarity patches in the SOT/SP data are consistent 
with the MHD simulation (see Fig.~\ref{skt_pol}). The radial distribution of opposite 
polarity flux $B_{\rm{z}}^{\rm{opp}}$ as a fraction of the net flux \textbar$B_{\rm{z}}$\textbar\, 
is plotted in Fig.~\ref{opp_flux}. In the SOT/SP observations, opposite polarity flux appears even 
in the inner penumbra at $\log \tau= 0$, and reaches a fraction of up to $\sim55\%$ 
of the total flux at the outer boundary of the penumbra. The MHD simulation shows
$\sim47\%$ flux with opposite polarity at the outer boundary of the penumbra,
but starting at a larger $r/R_{\rm{spot}}$ value. 
At $\log \tau= -0.9$ the SOT/SP data have only $\sim15\%$ flux with 
opposite polarity at the outer boundary of the penumbra, whereas it amounts to 
$\sim30\%$ in the MHD simulation. Approximately $8\%$ of the flux has the 
opposite polarity in the outer penumbra at $\log \tau= -2.5$ in 
both the SOT/SP data and the MHD simulation. When the MHD simulation is degraded to 
$1.\arcsec0$ only a small fraction of the opposite polarity flux survives (see panel (c)
of Fig.~\ref{opp_flux}). With degraded spatial resolution a higher number of pixels with
negative $(\Delta B/\Delta \log \tau)$ is found in the penumbra compared to that in the native resolution of 
the MHD simulation (see panel (d) of Fig.~\ref{opp_flux}). 
In the umbra not a single pixel is found with negative values of $(\Delta B/\Delta \log \tau)$
in the degraded spatial resolution data but almost $\sim30\%$ of the pixels have 
negative values of $(\Delta B/\Delta \log \tau)$ in the native resolution presumably corresponding to 
umbral-dots \citep{Schuessler_2006,Riethmueller_2008a,Riethmueller_2013}. In summary, the low spatial resolution hides
opposite polarity patches in the sunspot. However, it causes many pixels with $(\Delta B/\Delta \log \tau)<0$,
hence, it also produces $(\Delta B/\Delta \log \tau)<0$ in the azimuthal averages.

\section{Discussion} \label{4_diss}

We determined the vertical gradient of the magnetic field strength throughout 
sunspots. From the VTT/TIP-2 observations we found that \textit{B} increases at a rate of approximately
1.3\,G\,km$^{-1}$ with geometrical depth above $\log \tau= 0.0$ in the umbra. 
In the SOT/SP data we found that $B$ increases with depth at an average rate of $\sim1.4$\,G\,km$^{-1}$ 
in the umbra. Here we are presenting the gradient in terms of geometrical depth just to facilitate comparisons with earlier studies.  
Similar values have been found by \citet{Borrero_2011} and \citet{Tiwari_2015} based on Hinode SOT/SP data and 
by \citet{Sanchez_cuberes_2005} from VTT/TIP data. \citet{Mathew_2003} report values around 
4\,G\,km$^{-1}$ and \citet{Westendrop_2001a} found 2.5\,G\,km$^{-1}$. 
We found that $B$ increases rapidly with depth in the lower atmosphere 
while the gradient becomes flatter in the upper atmosphere above $\log \tau= -0.9$ 
in the SOT/SP data. This might explain the difference to that found by \citet{Mathew_2003}, who
obtained the gradient from the inversion of the \ionn{Fe}{i}\,1.56\,$\mu$m lines
that form very deeply in the photosphere. 

\begin{figure}
\centering
   \includegraphics[width=0.45\textwidth]{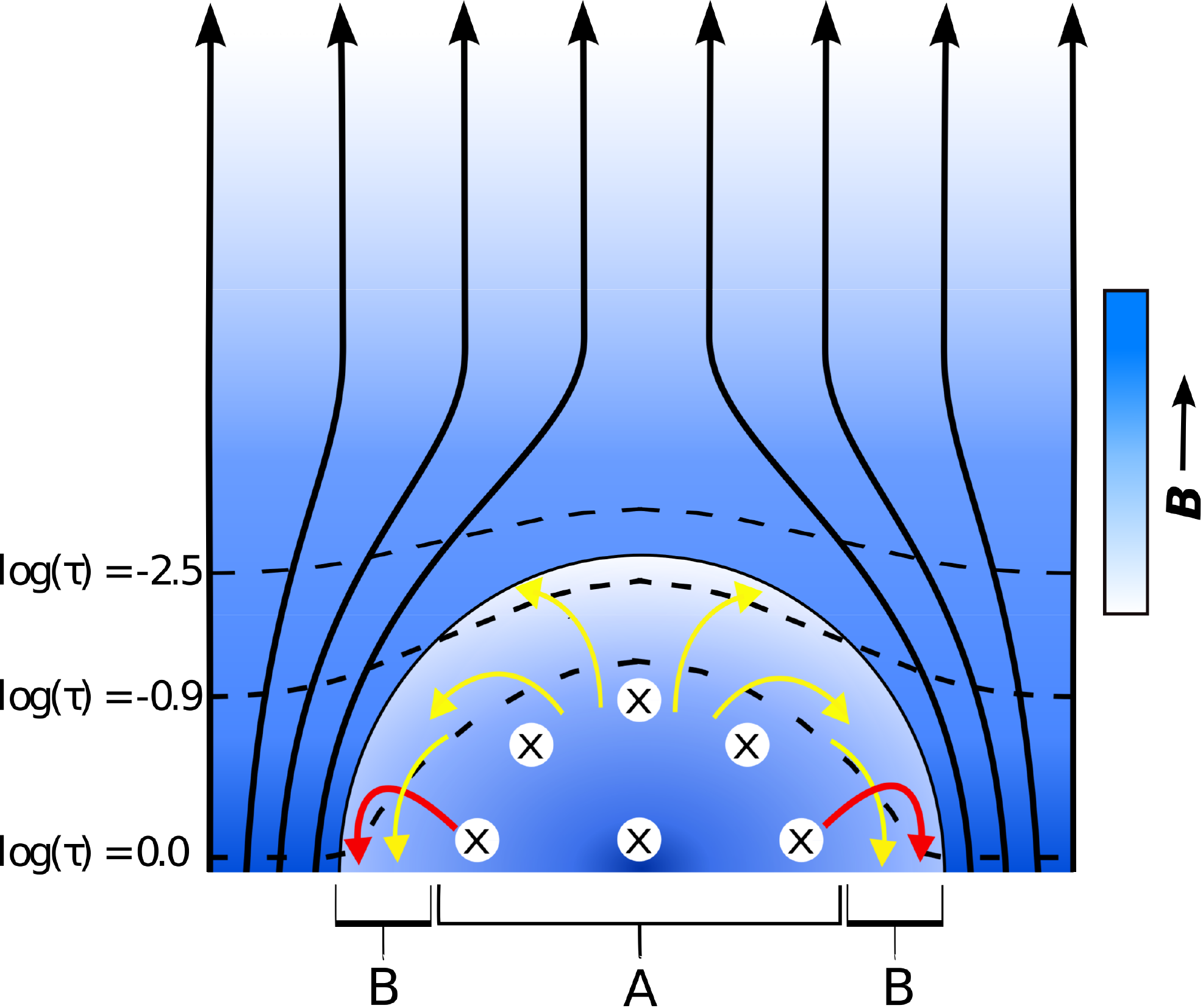}
      \caption{Schematic view of the vertical structure of the magnetic 
               field in penumbral filaments following \citet{Zakharov_2008}.
               Depicted is a plane perpendicular to the filament. The filament is represented  
               by the semi-circular area. \textit{Crosses} in circular \textit{white} areas indicate 
               the direction of the magnetic field in the penumbral filament, which is 
               perpendicular to the plane of the drawing. Arrows in \textit{yellow} indicate the overturning
               convection  within the filament and \textit{red} arrows depict the field reversal 
               at the edges of the filaments, which is caused by the convective flow through 
               advection of the filament's field. Only the projected component of the field,
               which is directed mainly into the page even for the \textit{red} arrows, is plotted. 
               The background magnetic field (spines) is indicated by nearly vertical \textit{black} 
               arrows. \textit{Dashed} lines show different $\log \tau$ levels.\,\textit{Lighter} 
               and \textit{darker shades} of \textit{blue} correspond to weaker and stronger $B$, respectively.
               }
\label{fil_car}
\end{figure}

The vertical gradient of the magnetic field strength in the penumbra, reported
in the present work, displays large fluctuations, changing its sign even on small
scales. In the gradient maps of the penumbra derived from the SOT/SP observations,
$(\Delta B/\Delta \log \tau)$ is always positive in spines. The gradient in the penumbral filaments
has a more complex structure: in the lower layer of the inner half of the filaments (i.e., the part of a filament closer 
to the umbra) has a positive $(\Delta B/\Delta \log \tau)$ which is surrounded by a negative $(\Delta B/\Delta \log \tau)$
at the sides. In the upper layer of the inner half part of the filaments it has only a negative $(\Delta B/\Delta \log \tau)$ 
that is strongest along the central ridge of the filaments. The outer half 
of the filament has a positive $(\Delta B/\Delta \log \tau)$ both in the lower and upper layers. The positive $(\Delta B/\Delta \log \tau)$ 
in the tail can be explained by the fact that the horizontal field of the filament, which is returning to the surface 
is stronger than the spine's field covering the filaments from the top \citep{Tiwari_2013}.
The decrease in the number and strength of spines with $r/R_{\rm{spot}}$ probably also contributes. 
The structure of the gradient in the body of filaments close to the head can be explained by a penumbral model presented in 
Fig.~\ref{fil_car} (adopted from \citet{Zakharov_2008}). The vertical structure 
of a penumbral filament is assumed to be a semi-circular horizontal cylinder 
containing nearly horizontal magnetic field. The filament is surrounded by a more vertical 
background magnetic field (spines) on the sides, which closes above the filament. The background field forms spines 
with $B$ gradually decreasing with height. This model assumes that the semicircular filament 
has its maximum magnetic field strength at its center, which gradually decreases towards the boundary
between the spines and the filament.
The $\log \tau= 0.0$ level
samples the whole width of the filament (shown by region B+A+B in Fig.~\ref{fil_car}),
whereas the $\log \tau= -0.9$ level samples only the central part of the filament's 
width (region A). Finally, the $\log \tau= -2.5$ level sees field only from the spines.
In region A the field strength decreases between $\log \tau= 0.0$ and $\log \tau= -0.9$,
hence, $(\Delta B/\Delta \log \tau)_{0.0,-0.9}$ is positive. In region B $(\Delta B/\Delta \log \tau)_{0.0,-0.9}$ is negative 
because the $\log \tau= 0.0$ level probes the filament, whereas the $\log \tau= -0.9$ level 
samples the stronger field from the spines.  
$(\Delta B/\Delta \log \tau)_{-0.9,-2.5}<0$ in region A because the $\log \tau= -0.9$ level is still 
within the filament's body whereas the $\log \tau= -2.5$ level is above the filament where 
the stronger field from the spines closes. In summary, the expanding and closing magnetic field from 
the spines above the filament causes the negative $(\Delta B/\Delta \log \tau)$ in the inner part of penumbral 
filaments.

In the MHD simulation $(\Delta B/\Delta \log \tau)$ is negative in the inner part of filaments
in both layers. $(\Delta B/\Delta \log \tau)_{0.0,-0.9}$ is positive in the tail of filaments,
whereas $(\Delta B/\Delta \log \tau)_{-0.9,-2.5}$ is negative there. The gradient maps thus
reveals that there are differences in the details of magnetic field structure of the penumbral 
filaments between the MHD simulation and the observations.

After producing azimuthal averages, our results turn out to be rather 
different from results published previously \citep{Westendrop_2001a,Mathew_2003,
Sanchez_cuberes_2005,Balthasar_2008,Borrero_2011}. In the VTT/TIP-2 data, where we 
assume that $B$ varies linearly with $\log \tau$, we found that \textit{B} decreases with depth 
from the umbra-penumbra boundary to $r\simeq 0.75 R_{\rm{spot}}$.  
Between $r\simeq 0.75 R_{\rm{spot}}$ and $r/R_{\rm{spot}}=1.0$ \textit{B} 
increases with depth. The SOT/SP observations exhibit a similar profile of $(\Delta B/\Delta \log \tau)$ 
in the lower layers. In the upper atmosphere $B$ increases on average with depth at all 
radial distances in the penumbra. $(\Delta B/\Delta \log \tau)_{0.0,-0.9}$ estimated 
from the MHD simulation shows good qualitative similarities with the SOT/SP data  
in the lower layers, although, $(\Delta B/\Delta \log \tau)_{0.0,-0.9}$ in the 
MHD simulation is mostly close to zero (slightly negative) in the inner penumbra. 
The average radial profiles of $(\Delta B/\Delta \log \tau)$, found in the 
present study with VTT/TIP-2 and SOT/SP observations disagree 
with earlier observational studies. \citet{Westendrop_2001a} and \citet{Borrero_2011} found that 
\textit{B} increases with depth in the inner half radius of the sunspot. In the 
outer half radius the field decreases with depth. In contrast, 
\citet{Mathew_2003} and \citet{Sanchez_cuberes_2005} found that \textit{B} 
increases with depth everywhere within the sunspot's visible boundary.
An observation of a reversed gradient in the vertical magnetic field component
in the penumbra has been reported by \citet{Balthasar_2013}, based on difference between 
the magnetic field inferred from the \ionn{Fe}{i}\,10783\,\AA\, and \ionn{Si}{i}\,10786\,\AA\, lines.

The expansion of the sunspot's magnetic field due to the decreasing gas pressure, $(\partial P/\partial d)>0$,
should always produce a positive $(\Delta B/\Delta \log \tau)$ on a global 
scale. This is in seeming contradiction with the azimuthally averaged radial profiles of $(\Delta B/\Delta \log \tau)_{0.0,-0.9}$
found in the SOT/SP observations and the MHD simulation where $(\Delta B/\Delta \log \tau)_{0.0,-0.9}$ is negative 
in the inner penumbra. Highly corrugated optical depth surfaces, sampling  
the atmosphere at different geometrical depths in different parts of the penumbra, can cause the observed negative 
gradient in the azimuthal averages. In the penumbral 
filaments, the $\log \tau$ surfaces are elevated compared to the spines (see Fig~\ref{MHD_ver} and Fig~\ref{fil_car}).
In the spines the magnetic field strength changes gradually.
in contrast to the filament exhibiting an abrupt change in the magnetic field strength 
(from filament to the overlying field of the expanding spines) between two $\log \tau$ surfaces. 
The scenario described above leads to a negative $(\Delta B/\Delta \log \tau)_{0.0,-0.9}$ with higher 
magnitude in and above the filaments compared to the positive $(\Delta B/\Delta \log \tau)_{0.0,-0.9}$ in the spines. 
The negative $(\Delta B/\Delta \log \tau)_{0.0,-0.9}$
is only seen around the heads of the filaments. A higher density of filament heads at a given $r/R_{\rm{spot}}$ leads to a higher probability 
of negative values for the azimuthal averages of $(\Delta B/\Delta \log \tau)_{0.0,-0.9}$. This is the case in the inner
penumbra, compared to the middle and outer penumbra.

The statement that the highly corrugated $\log \tau$ surfaces cause the negative gradients in 
the azimuthal averages is supported by the MHD simulation. The azimuthally averaged 
gradient with respect to the geometrical scale, $\partial B/\partial d$, from the MHD simulation does not show negative values in the inner penumbra     
above $\log \tau = 0$. When $(\Delta B/\Delta \log \tau)$ is examined then only the MHD simulations show a negative or close to zero vertical gradient of 
the  magnetic field in the inner penumbra.

An underestimation of $B$ in deeper layers could also contribute to the observed
negative gradient in the inner penumbra. Underestimation of $B$ can be caused by a cancellation of polarized 
signal stemming from unresolved opposite polarity areas. The presence of opposite polarity 
patches in sunspot penumbrae has been observed 
by \citet{Scharmer_2013}, \citet{Ruiz_2013,Franz_2013}, \citet{Tiwari_2013} and \citet{Vannoort_2013}. 
\citet{Ruiz_2013} and \citet{Franz_2013} have found the opposite polarity patches in the middle and outer penumbra 
in SOT/SP observations while \citet{Scharmer_2013} found thinner and elongated lanes of 
opposite polarity everywhere in the penumbra in higher resolution SST/CRISP observations. 
Recent 3D MHD simulations of a sunspot by \citet{Rempel_2012}; cf. \citet{Rempel_2009a} 
also exhibit opposite polarity patches in the penumbra. These patches of 
opposite polarities are thought to be produced by advection of the field by the
overturning convection in penumbrae. In recent years many observational 
evidences for overturning convection in penumbrae has been reported 
\citep{Joshi_2011,Scharmer_2011,Tiwari_2013,Esteban_2015}.

Our analysis of opposite polarity patches in the SOT/SP observations and the MHD simulation indicates that flux 
associated with opposite polarity patches increases from the inner to the outer penumbra 
\citep[see also Fig.~13 of][]{Rempel_2012}. Both the MHD simulation and the SOT/SP observations 
reveals the presence of opposite polarity patches even in the higher part of the atmosphere, i.e., up to 
$\log \tau= -2.3$. The distribution of opposite polarity patches in the penumbra 
observed with SOT/SP is qualitatively consistent with that of the MHD 
simulation (see Fig.~\ref{skt_pol} and Fig.~\ref{opp_flux}), 
although there is a quantitative difference in the flux associated with opposite polarity patches
in the SOT/SP observations and the MHD simulation. In the upper layers opposite polarity patches 
are restricted to the middle and outer penumbra both in the SOT/SP observations and the MHD simulation. 
These opposite polarity patches are of two kinds: narrower and elongated lanes of opposite polarity at the edges 
of the inner part of penumbral filaments (see Fig.~\ref{fil_car}) at the locations of downflow lanes which are the 
lateral components of overturning convection \citep{Scharmer_2013,Tiwari_2013}. Patches found 
at the tails of the penumbral filaments are bigger in size \citep{Franz_2013,Ruiz_2013,
Scharmer_2013,Tiwari_2013,Vannoort_2013} and are associated with downflows at the tails.

Our experiment, estimating $(\Delta B/\Delta \log \tau)_{0.0,-2.3}$ from the degraded 
spatial resolution ($1.\arcsec0$) MHD simulation shows that at this resolution
opposite polarity patches disappear almost completely. At the same time a significant 
increase in the number of pixels with negative $(\Delta B/\Delta \log \tau)_{0.0,-2.3}$ occurs in 
the penumbra. This behavior reverses in the umbra where in the native resolution $\sim30 \%$ of the 
pixels have negative $(\Delta B/\Delta \log \tau)_{0.0,-2.3}$ which disappear completely in the degraded resolution data.
As discussed above, we have found more flux with opposite polarity in the lower layers compared 
to upper layers, thus lower spatial resolution causes more cancellation in 
the polarized signal from deeper layers and produces negative $(\Delta B/\Delta \log \tau)_{0.0,-2.3}$. 

As far as the SOT/SP observations are concerned, a significant amount of flux is associated with opposite 
polarity patches. But it is still possible that narrow elongated opposite 
polarity lanes at the edges of the penumbral filaments are not completely resolved. In this case  
it is possible that some fraction of the negative gradient originates from unresolved 
opposite polarity patches. However, the radial distribution of opposite polarity patches and number of pixels with 
negative values of $(\Delta B/\Delta \log \tau)_{0.0,-2.3}$ are rather different. This suggests that
the main cause of the observed negative $(\Delta B/\Delta \log \tau)_{0.0,-2.3}$ in the azimuthal averages 
are the highly corrugated iso-$\tau$ surfaces.

The presence of a magnetic canopy outside the visible boundary of the sunspot 
\citep{Solanki_1992b,Lites_1993,Solanki_1994,Adams_1994,Skumanich_1994,
Keppens_1996,Rueedi_1998,Solanki_1999}, 
with the canopy base rising with distance from the sunspot, can be concluded from the SOT/SP 
observations. This is also consistent with the results from the MHD simulation and in line with the results of \citet{Tiwari_2015}.
The vertical gradient of $B$ turns out to be a useful quantity for identifying 
or detecting magnetic canopies.

An important result of this study is that on large scales the magnetic canopy only 
exists at and outside the visible boundary of the sunspot ($r/R_{\rm{spot}}>0.9$). 
We do not see a canopy structure in the middle 
penumbra caused by the expansion of field from the umbra or 
the inner penumbra as reported by \citet{Westendrop_2001a} and \citet{Borrero_2011}. 
Instead, we found local patches in the penumbra where the magnetic field 
strength decreases with optical depth caused by the stronger field in the spines closing 
above the filaments (see Fig.~\ref{fil_car}). The corrugation effect 
of the $\tau$ surfaces leads to negative $(\Delta B/\Delta \log \tau)$ in the inner penumbra in the lower atmosphere 
in the azimuthal average. In addition to the corrugation of the $\tau$ surfaces,
the effect of unresolved, deep-lying opposite polarity patches can also lead to negative 
$(\Delta B/\Delta \log \tau)$ patches in the penumbra.

\begin{acknowledgements}
The German Vacuum Tower Telescope is operated by the Kiepenheuer-Institut f\"{u}r
Sonnenphysik at the Spanish Observatorio del Teide of 
the Instituto de Astrof\'{\i}sica de Canarias (IAC).
Hinode is a Japanese mission developed and launched by
ISAS/JAXA, with NAOJ as domestic partner and NASA and STFC (UK) as
international partners. It is operated by these agencies in co-operation with
ESA and NSC (Norway). We thank M. Rempel for providing the MHD simulations. JJ acknowledges a PhD fellowship of the
International Max Planck Research School on Physical Processes in the Solar
System and Beyond (IMPRS). This work was partly supported by the BK21 plus
program through the National Research Foundation (NRF) funded by the Ministry of Education of Korea.
SKT is supported by an appointment to the NASA Postdoctoral Program at the NASA Marshall Space Flight 
Center, administered by Universities Space Research Association under contract with NASA.
\end{acknowledgements}

\bibliographystyle{aa} 
\bibliography{joshi_thesis}
\end{document}